\documentclass{article}

\usepackage{vmargin,amssymb,amsmath,epsfig}

\newcommand{\eqsp}{\, = \,}  
\newcommand{\rar}{\, \rightarrow \,}
\newcommand{\beq}{\begin{equation}}
\newcommand{\eeq}{\end{equation}}
\newcommand{\la}{\langle}
\newcommand{\ra}{\rangle}
\newcommand{\ep}{\epsilon}
\newcommand{\cX}{{\cal X}}
\newcommand{\Ga}{\Gamma}

\newcommand{\dupp}{\delta u^{++}}
\newcommand{\cS}{\cal S}
\newcommand{\nFour}{${\cal N} \, = \, 4 \ $}

\newcommand{\cY}{{\cal Y}}
\newcommand{\cZ}{{\cal Z}}
\newcommand{\mS}{\mathbb{S}}

\newcommand{\fq}{\mathbf{q}}
\newcommand{\fs}{\mathbf{s}}
\newcommand{\fr}{\mathbf{r}}
\newcommand{\ft}{\mathbf{t}}
\newcommand{\fv}{\mathbf{v}}
\newcommand{\fw}{\mathbf{w}}

\begin{document}

\thispagestyle{empty}


\phantom{w}

\vskip 4.5 cm

\begin{center}

{\huge \textbf{One-loop five-point gluing analytically \\[1 mm] evaluated}}

\vskip 1.5 cm

{\large B.~Eden, M.~Gottwald$^a$}
\vskip 1 cm

$^a$ Institut f\"ur Mathematik und Physik, Humboldt-Universit\"at zu Berlin, \\ Zum gro{\ss}en Windkanal 2, 12489 Berlin, Germany \\[2 mm]
\vskip 1 cm

e-mail: gottwalm@physik.hu-berlin.de

\end{center}

\vskip 3 cm

\textbf{Abstract:} \\[-2 mm]

The one-loop five-point function of stress tensor multiplets in \nFour super Yang-Mills theory in four dimensions has previously been studied by integrability methods. The finite-size viz \emph{gluing} corrections defining it were originally matched on the known field theory result --- establishing many features of the formalism on the way --- and later to a good extent also analytically evaluated. For the first time, we present a full analytic evaluation of all processes. The starting point is the re-summation of series of residues into Euler integrals. These are directly integrated where possible or otherwise recovered from intersection theory for generalised hypergeometric functions. 

\newpage

\section{Introduction}

The \nFour super Yang-Mills theory in four dimensions is an example of an exactly conformally invariant quantum field theory. Quantities of interest are the correlation functions of gauge invariant composite operators. With unbroken supersymmetry there are no non-trivial vacuum expectation values so that the first interesting object is the two-point function. It obeys a power law in which the exponent --- the anomalous dimension --- can be calculated from Feynman graphs, or from string data thanks to the AdS$_5$/CFT$_4$ correspondence \cite{123}. 

The computation of planar anomalous dimensions has been recast as finding the energy levels of an integrable system \cite{beiStau}. This approach has led to spectacular progress: for certain quantities, eg. the so-called \emph{scaling function} the weak- \cite{ES,BES} and strong-coupling \cite{Grisha} expansion can be derived to many orders in the 't Hooft coupling, and numerical interpolation is possible \cite{Klebanov}. Similar success became possible for anomalous dimensions of local operators via the \emph{quantum spectral curve} \cite{QSC}.

Yet, for a long time higher-point functions remained hard to address using these methods. The introduction of the \emph{hexagon operator} meant a break-through wrt. the three-point problem \cite{BKV}. Finally, it was noticed in \cite{cushions,shotaThiago1} that higher-point functions can be evaluated by tilings with hexagon patches. There is a price to pay: the tiles have to be glued together by the exchange of virtual particles, and such gluing processes are hard to evaluate. 

In the context of the AdS$_5$/CFT$_4$ correspondence, higher-point functions of BPS operators --- so operators with vanishing anomalous dimension --- have received a lot of attention \cite{usBianchiGleb} because their exact strong coupling behaviour is captured by supergravity. The most advanced \emph{integrability} results also belong here: considering operators built out of very many scalar fields of the theory all-loops results can be derived, even including non-planar corrections \cite{octagons}.

Beyond four points nothing similar has been achieved, and quite generally without the help of any semi-classical approximation as the one described in the last paragraph we find that the study of gluing corrections is just as hard as --- or perhaps even worse than --- the direct approach to the weak coupling expansion by Feynman graphs. Indeed, the gluing corrections are closely related \cite{shotaThiago1} to so-called Mellin-Barnes \cite{MB} representations of Feynman diagrams.

\begin{figure}[h]
\begin{center}
\includegraphics[width=5.5 cm]{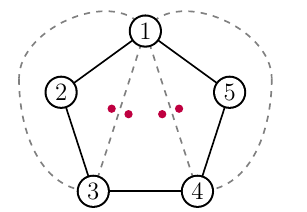}
\caption{Gluing three adjacent tiles by two virtual exchanges; the conjugate pairs of bound states are marked by the dots on the two sides of the glued edges.}
\end{center}
\label{figureGlu}
\end{figure}

In this article we report on progress in our quest for analytic methods for the evaluation of gluings. In particular, we study the situation displayed in Figure \ref{figureGlu} in which three adjacent tiles of the planar five-point stress tensor correlator are glued by one virtual particle passing each common edge.
\vskip- 0.475 cm
Let us label the top centre point as 1, then counter-clockwise around Figure \ref{figureGlu}. This is a five-point one-loop contribution; one could imagine a semi-circular Yang-Mills line passing from the upper left edge through the middle of the figure to the upper right one. This process has been evaluated in \cite{shotaThiago2} by matching a truncated residue calculation on the quantum field theory result \cite{drukkerPlefka}.  In \cite{usFivePoints} we have been able to directly integrate enough of the pieces of the entire process to guess a function space sufficient to capture most of the individual processes, and to match the sum of all parts onto these functions. In the present article we finally present a fully analytic evaluation. 

The article is organised as follows:
\begin{itemize}
\item We provide a structural overview of the problem in Section \ref{setUpProb}.
\item In Section \ref{hardest} we use the top complicated part of any individual process to illustrate how the re-summation of residues results into Euler integrals with up to four parameters. Some of these can be directly integrated, others are multi-quadratic in the integration parameters, wrt. one variable even cubic. We scale out one or two parameters by variable redefinitions and submit the resulting biquadratic problem to intersection theory for generalised hypergeometric functions \cite{inter1,inter2}. The result is rescaled and re-imported, the remaining integrals now have multi-linear denominators.
\item In Section \ref{bivariate} we explain the steps of the intersection theory computations on the example of Section \ref{hardest}.
\item In the following sections all other series of residues are studied and the results put together.
\item We end on some conclusions.
\item Appendix A contains an alternative attempt on the computation of Section \ref{hardest}, starting from yet another form of the bound state scattering matrix. We comment on some aspects of the intersection calculation in Appendix B. Appendix C lists and explains the various ancillary files.
\end{itemize}

\section{Setup} \label{setUpProb}

In the integrability approach to the spectrum problem the physical excitations carry an undotted and a dotted super-index that transform under a central extension of the supergroup $PSU(2|2)_L \otimes PSU(2|2)_R$. Scattering takes place on a left and a right super spin chain under a product of two $S$ operators as given eg. in \cite{beisertPsu22}. The hexagon \cite{BKV} employed in three-point functions has a diagonal $PSU(2|2)$ symmetry left after moving the three space time (and internal space) points at its corners to $0, \, 1, \, \infty$, respectively. It is constructed out of the bricks of the integrability approach to the spectrum problem, although scattering now takes place only on the left or the right chain --- it does not matter which one --- while the result is contracted in a certain way onto the other chain. For the process in the Figure \ref{figureGlu} this implies diagonal scattering, cf. \cite{shotaThiago2,usFivePoints}. 

Next, all particles on a hexagon can be moved to one edge by crossing/mirror transformations \cite{BKV}. The scattering matrix of the spectrum is written in terms of the Zhukowski function $x(u,g)$ obyeing
\beq
x + \frac{1}{x} \eqsp \frac{u}{2 \, g}.
\eeq
Here $u$ is a rapidity of the particle in question and $g \eqsp \sqrt{g_{YM}^2 N/(4 \pi^2)}$ is the 't Hooft coupling. In physical kinematics and at weak coupling we choose the branch satisfying $x \eqsp u/(2 g) + \ldots$. Next, the particle rapidities are typically shifted into the complex plane:
\beq
x^\pm \eqsp x \left(u \pm \frac{i \, K}{2}\right) \, , \qquad K \, \in \, \mathbb{N} .
\eeq
A fundamental particle has $K \eqsp 1$, but there are bound states built out of $K$ particles that are moving as a compound with one common rapidity $u$. The transformations
\begin{eqnarray}
1 \gamma & : & x^+ \, \rightarrow \, \frac{1}{x^+} \, , \quad x^- \rar x^- \, , \nonumber \\ 
2 \gamma & : & x^\pm \, \rightarrow \, \frac{1}{x^\pm} \, , \nonumber \\
3 \gamma & : & x^+ \, \rightarrow \, x^+, \ x^- \, \rightarrow \, \frac{1}{x^-} \, ,  \\
4 \gamma & : & x^\pm \, \rightarrow \, x^\pm \, . \nonumber
\end{eqnarray}
($5 \gamma$ is $1 \gamma$ etc.) can be used to move all particles to $0 \gamma$ edge (we choose one of the locations of the outer operators), so counting anti-clockwise a particle on the $(n+1)$-th edge is brought to the first edge by an $n \, \gamma$ transform. We then scatter and contract as described; for details we would ask the interested reader to consult \cite{BKV}. In particular, for the central hexagon in Figure \ref{figureGlu} we assign $0 \gamma$ to the bottom right point in the figure (point 4) so that the left bound state is moved there by a $3 \gamma$ transform and the right one by $1 \gamma$. Importantly, the two magnons arrive on the same chain in the original ordering, so $X(u^{3 \gamma}) \, Y(v^{1 \gamma})$. 

The scattering has a matrix part given by the \emph{bound state scattering matrix} derived in \cite{glebBound}. Beyond it there is the scalar factor
\beq
h(u^{3 \gamma}, \, v^{1 \gamma}) \eqsp \Sigma(u^{1 \gamma}, v^{1 \gamma})  \eqsp \frac{\Gamma[1 + \frac{K}{2} + i \, u]}{\Gamma[1 + \frac{K}{2} - i \, u]} \, \frac{\Gamma[1 + \frac{L}{2} - i \, v]}{\Gamma[1 + \frac{L}{2} + i \, v]} \, \frac{\Gamma[1 + \frac{K+L}{2} - i (u-v)]}{\Gamma[1 + \frac{K+L}{2} + i (u-v)]} \, + O(g) \, . \label{sig11}
\eeq
where $\Sigma$ in that $\gamma$ rotation equals the \emph{improved BES dressing phase} in \emph{mirror/mirror} kinematics \cite{BES,glebMirror}. We have associated the bound state length $K$ to the first particle with rapidity $u$ and the bound state length $L$ with the second one having rapidity $v$. We will also associate $x^\mp$ to the first particle and $y^\mp$ to the second where convenient.

We will need bound states in the \emph{antisymmetric representation}: due to the statitstics they can be compounds of many fermions but only of one or two bosons: recall that the individual particle here lives on, say, the left chain and hence has a single super index $D \eqsp (d, \delta)$ where both, the bosonic part $d$ and the fermionic one $\delta$ are two component indices. We can clearly antisymmetrise many fermions, but maximally two distinct bosons. There are thus four types of bound states:
\beq
(\psi^1)^{K-k-1} (\psi^2)^k \phi^d \, , \qquad (\psi^1)^{K-k} (\psi^2)^k \, , \qquad (\psi^1)^{K-k-1} (\psi^2)^{k-1} \phi^1 \phi^2 \, . \label{sl2BoundBits}
\eeq
Hence the total number of constituents is $K$ whereas the magnetic quantum number $k$ counts the number of up spins, say, within the total number of spinor indices. In a scattering process, the bound state lengths $K, \, L$ of the two states will be conserved. However, the $sl(2)$ label is not. Using the short $|k,l\ra \eqsp |k \ra \otimes | l \ra$ for the tensor product of two states of the first type with scalar constituent $\phi^1$
\beq
\mathbb{S} \, |k,l\ra \, = \, \sum_{n=0}^{k+l} \, \cX^{k,l}_n \, |n,k+l-n\ra \label{opS11}
\eeq
where $n \, = \, k$ is diagonal, so we adopt the convention in which the state with bound state length $K$ is written on the left before and after scattering. For simplicity, let us index the type of state by the flavour of the scalars involved:
\beq
| 1 \ra, \, | 2 \ra , \, |-\ra, \, |12\ra
\eeq
Beyond $\cX$ other diagonal bound state $\mS$ matrix elements appear in our computation:
\begin{equation}
\begin{aligned}
&\mS \, |1,- \ra  \!&=&\;  \cY_{11} \, |1,-\ra \, , &&\qquad \mS \, |-,1\ra \!&=&\; \, \cY_{22} \, |-,1\ra \, ,\\
&\mS \, |1,12 \ra \!&=&\; \cY_{33} \, |1,12\ra \, , &&\qquad \mS \, |12,1 \ra\! &=&\; \cY_{44} \, |12,1\ra \, ,\\
&\mS \, |-,- \ra  \!&=&\;  \cZ_{11} \, |-,-\ra \, , &&\qquad \mS \, |-,12 \ra  \!&=&\;  \cZ_{22} \, |-,12\ra \, ,\\
&\mS \, |12,- \ra  \!&=&\; \, \cZ_{33} \, |12,-\ra \, , &&\qquad \mS \, |12,12 \ra \!&=&\; \, \cZ_{44} \, |12,12\ra,\\
&\mS \, |1,2 \ra  \!&=&\;  \cZ_{55} \, |1,2\ra \, , &&\qquad \mS \, |2,1 \ra \!&=&\; \, \cZ_{66} \, |2,1\ra \, . 
\end{aligned}
\label{defY}
\end{equation}
There is a second copy $\tilde \cX$ algebraically equal to $\cX$ for the scattering of the states $|2,2\ra$ and in the same way $\tilde \cY$ (identical to $\cY$) is defined as in \eqref{defY} but for single scalars of flavour~two. In the original work \cite{glebBound} $\cY$ and $\cZ$ are expressed by a linear combination of $\cX$ matrices with shifted indices. However, their coefficient matrices are hardly suited to our purpose of re-summing series of residues and subsequent integration due to their complicated polynomial denominators, which were only eliminated in the later work \cite{ETH}. The latter publication comes with a notebook on which we build in the present article. Let us emphasise that the original and the later writing of the $\cY, \, \cZ$ elements, respectively, are algebraically equal, although this is not at all easy to realise because of the square root nature of the $x^\mp$ functions involved.

The $\cX$ matrix is defined as follows:
\begin{eqnarray}
&& \cX_n^{k,l} \, = \, D \ \frac{\prod_{j=1}^n (K-j) \, \prod_{j=1}^{k+l-n} (L-j)}{\prod_{j=1}^k (K-j) \, \prod_{j=1}^l (L-j) \, \prod_{j=1}^{k+l} (-i \, \delta u + \frac{K+L}{2} - j)} \, * \nonumber \\
&& \sum_{m=0}^k \left(\begin{array}{c} k \\ k-m \end{array} \right) \left(\begin{array}{c} l \\ n-m \end{array} \right) \, \prod_{j=1}^m c^+_j \, \prod_{j=1-m}^{l-n} c^-_j \, \prod_{j=1}^{k-m} d_{k-j+2} \, \prod_{j=1}^{n-m} \tilde d_{k+l-m-j+2}  \, , \nonumber \label{defX}
\end{eqnarray}
\vskip -0.25 cm
\beq
c^\pm_j = -i \, \delta u \, \pm \frac{K-L}{2} - j + 1 \, , \quad d_j =\frac{K + 1 - j}{2} \, , \quad \tilde d_j = \frac{L + 1 - j}{2} \,  , 
\eeq
where $\delta u \, = \, u - v$ and
\beq
D \, = \, \frac{x^- -y^+}{x^+ - y^-} \sqrt{\frac{x^+}{x^-}} \sqrt{\frac{y^-}{y^+}} \, .
\eeq
Under the $3\gamma, 1\gamma$ transformation $x^- \, \rightarrow \, 1/x^-, \, y^+ \, \rightarrow \, 1/y^+$ and to
lowest order in $g$
\beq
D \, = \, - \frac{u^- -v^+}{u^+ - v^-} \frac{\sqrt{u^+ u^-} \sqrt{v^+ v^-}}{u^- v^+} \, . \label{dWrong}
\eeq
Neither of the dressing factor \eqref{sig11}, the matrix part \eqref{defX} of the scattering  \eqref{opS11}, or a measure factor\footnote{No additional crossing or mirror transforms are applied to $M$.}
\beq
M \eqsp \frac{g^4 \, K \, L}{4 \, \pi^2 \, (u^- \, u^+ \, v^- \, v^+)^2} + \ldots \label{defM}
\eeq 
display square roots of the rapidities. Do the branch cuts of $D$ in the $3 \gamma \, 1 \gamma$ kinematics thus spoil our chances of picking residues? The last building block are weight factors related to the kinematics, to be defined in \eqref{defW} below. Unfortunately, those will not help either. What is missing are braiding factors involving the momentum. Defining $U \eqsp e^{i p/2} \eqsp \sqrt{x^+/ x^-}$ we must introduce
\beq
\frac{U_1}{U_2} \ \stackrel{3 \gamma, \, 1 \gamma}{\longrightarrow} \ \frac{\sqrt{u^- u^+} \sqrt{v^- v^+}}{g^2}
\eeq
into the $\cX$ scattering \eqref{opS11}, and the \emph{same} for the other copy of $\cX$, whereby there is no obvious link to the flavour of the scalar constituents. This will at the same time combine with the square roots  in \eqref{dWrong}, lower the power of $g$ in \eqref{defM} to $g^2$ as expected for a one-loop contribution, and eventually yield a result of logarithm weight two. Appropriate braiding factors for the other diagonal elements of the bound state $S$ matrix are
\beq
\cY_{11} , \,  \cY_{33} \ : \ 1/U_2 \, , \qquad \cY_{22}, \,  \cY_{44}  \ : \ U_1 \, , \qquad \cZ_{11}, \, \cZ_{22}, \, \cZ_{33}, \, \cZ_{44} \ : \ 1 \, ,
\eeq 
while there are two possible choices for $\cZ_{55}, \, \cZ_{66}$ namely to decorate both by $U_1 \, U_2$ or both by $1/(U_1 \, U_2)$. These were called the $+, \, -$ grading in \cite{shotaThiago2} and an \emph{averaging prescription} was chosen in which both gradings are added with weight $1/2$. In fact, either of the two gradings would be sufficient on its own \cite{usFivePoints}. 

As we had mentioned, in \cite{BKV} the space time and internal space coordinates were moved to the standard position $0, \, 1, \, \infty$ in order design a formalism for evaluating only structure constants rather than full three-point functions. Here we do want to re-instate the dependence on kinematic invariants as we are dealing with a non-trivial function of conformal cross ratios. In \cite{shotaThiago1,shotaThiago2} it is elaborated how to rotate back hexagon operators to more general kinematics. Following \cite{shotaThiago2} we restrict the full five-point problem to kinematics in a two-dimensional plane. Hence on both, Minkowski and internal space the fifth normally independent cross ratio will be a function of the other four. The outcome is to associate the cross ratios
\beq
z_1 \, \bar z_1 \eqsp \frac{x_{14}^2 x_{23}^2}{x_{12}^2 x_{34}^2} \, , \qquad  (1-z_1) \, (1-\bar z_1) \eqsp \frac{x_{13}^2 x_{24}^2}{x_{12}^2 x_{34}^2} \label{cross1}
\eeq
with the virtual particle crossing the 13 edge. Cyclically shifting we obtain
\beq
z_2 \, \bar z_2 \eqsp \frac{x_{15}^2 x_{34}^2}{x_{13}^2 x_{45}^2} \, , \qquad  (1-z_2) \, (1-\bar z_2) \eqsp \frac{x_{14}^2 x_{35}^2}{x_{13}^2 x_{45}^2}
\eeq
for the second virtual particle, and similarly in internal space. The four space time variables will come into our residue computations via the weight factor
\beq
W \eqsp (z_1 \bar z_1)^{-i \, u} \, \left(\frac{z_1}{\bar z_1}\right)^{\frac{K}{2}-k} \ (z_2 \bar z_2)^{-i \, v} \, \left(\frac{z_2}{\bar z_2}\right)^{\frac{L}{2}-l} \, . \label{defW}
\eeq
when acting on a bound state $|-\ra$ without bosons or with both; otherwise there are shifts because of the modified range of the $k,\, l$ counters. There are also weight factors for the analogous internal space cross ratios $\alpha_1, \, \bar \alpha_1, \, \alpha_2, \, \bar \alpha_2$. These are much more trivial because there are no rapidity degrees of freedom in internal space. However, the rapidity part is replaced by the somewhat mysterious \emph{$J$ charge assignment}, which makes $\sqrt{z_j \, \bar z_j} / \sqrt{\alpha_j \, \bar \alpha_j}$ or its inverse appear \cite{shotaThiago2,usFivePoints}. Furthermore, the ratio-like part of the internal weights is similar but only causes multiplication or division by $\sqrt{\alpha_j / \bar \alpha_j}$ as there is at maximum a single scalar constituent of given flavour in each bound state. 

In conclusion the Figure \ref{figureGlu} needs to be evaluated for each distinct flavour choice for the bound states on the two central edges. In every case a complicated sum integral arises: there is a double integral over the two rapidities $u, \, v$ and a quadruple sum over the counters $K, \, k, \, L, \, l$. In the simplest case \eqref{opS11} featuring the $\cX$  matrix on its own, the integrand is a product of the measure, the weights, the scattering matrix (possibly supplemented by braiding factors), and the BES dressing phase \cite{BES}. In the more complicated cases, the single $\cX$ matrix will be replaced by a sum over $\cX$ matrices with shifted $k, \, l, \, n$ indices, each with some rational coefficient that we take from the output of the notebook published with \cite{ETH}. The notebook provided with \cite{ETH} decomposes the $\cZ$ elements of \cite{glebBound} wrt. to the spanning system
\beq
\cX \eqsp \{ \cX^{k,l}_n, \, \cX^{k-1,l}_n, \, \cX^{k,l-1}_{n-1}, \, \cX^{k,l}_n \} \, .
\label{eq:xBlocks}
\eeq
This is not unique because only any two index shifts of $\cX$ matrices are independent over rational functions of the rapidity difference $\delta_u$ and the various counters. The actual writing chosen in the notebook is one way of expressing the bound state scattering matrix without the complicated polynomial denominators of the $\cY$ and $\cZ$ elements that the original result \cite{glebBound} contains. 

To pick residues we close the contour of the $u$ integration over the positive half-plane, and that of the $v$ integration over the negative half-plane. This creates a power series in ascending powers of $z_1, \bar z_1$ but in descending powers of $z_2, \bar z_2$. To deal with all four variables on the same footing, and for convenience of notation in {\tt Mathematica}, we re-label as
\beq
\left\{z_1, \, \bar z_1, \, \frac{1}{z_2}, \, \frac{1}{\bar z_2} \right\} \rar \{ z, b, a, y \} \, .
\eeq

\subsubsection*{Characterising the residues}

After performing the mirror transformations, we follow the discussion of the pole structure from \cite{usFivePoints} always closing the contour for $u$ and $v$ in the upper and lower half-plane respectively. Furthermore, in order to avoid difficulties due to $\Gamma$-function in the phase $\Sigma$ or the matrix element, we first evaluate double poles (in either $u$ or $v$). Doing so, we can express the general contributions for any matrix element $\mathcal{M}$ like:
\begin{equation}
S(\mathcal{M})=\sum\int \tilde{\mu} \, W \, \Sigma^{KL}  \, (\hat R \, . \, \mathcal{X}) \, \text{d}u \, \text{d}v
\label{eq:contributionM}
\end{equation}
Here the factor $\tilde{\mu}$ contains the complete information about the pole structure from the measure and pre-factors from the matrix elements. Its explicit form is stated in Tab. \ref{tab:zvY}. The phase $\Sigma^{KL}$ is displayed in \eqref{sig11} and $\mathcal{X}$ are the two (for $\mathcal{Y}$'s) or four (in case of $\mathcal{Z}$'s) possible blocks described in \eqref{eq:xBlocks}. They are multiplied with a pre-factor $\hat R$ which does not contain any relevant poles.

\begin{table}[h]
\begin{center}
\begin{tabular}{c | c | c}
$\mathcal{M}$  & $\tilde{\mu}(u,\, v)$ & Contributions $S(\mathcal{M})^{u^-}$\& $S(\mathcal{M})^{v^+}$\\
\hline
\hline
$ $&$ $\\
$\mathcal{Y}_{11} $ &$\frac{1}{(u^-)^2 u^+ v^- v^+} $ &$S(\mathcal{Y}_{11} )^{u^-} $\\
$ $&$ $\\
$\mathcal{Y}_{22} $ &$\frac{1}{u^-u^+v^-(v^+)^2}$&$S(\mathcal{Y}_{22} )^{v^+} $\\
$ $&$ $\\
$\mathcal{Y}_{33} $ &$ \frac{(u^- - v^+)}{(u^-)^2u^+(v^+)^2} $&$S(\mathcal{Y}_{33} )^{u^-} $\& $S(\mathcal{Y}_{33} )^{v^+}$\\
$ $&$ $\\
$\mathcal{Y}_{44} $ &$ \frac{(u^- - v^+)}{(u^-)^2v^-(v^+)^2}$&$S(\mathcal{Y}_{44} )^{u^-} $\& $S(\mathcal{Y}_{44} )^{v^+}$\\
$ $&$ $\\
$\mathcal{Z}_{22} $&$\frac{1}{u^-u^+(v^+)^2} $&$S(\mathcal{Z}_{22} )^{u^-}$ \\
$ $&$ $\\
$\mathcal{Z}_{33} $&$\frac{1}{(u^-)^2v^-v^+}  $&$S(\mathcal{Z}_{33} )^{v^+} $\\
$ $&$ $\\
$\mathcal{Z}_{44} $&$\frac{(u^- - v^+)}{(u^-)^2 (v^+)^2}  $ &$S(\mathcal{Z}_{44} )^{u^-} $\& $S(\mathcal{Z}_{44} )^{v^+}$\\
\end{tabular}
\end{center}
\caption{The table provides an overview over the measure $ \tilde{\mu}$ and the type of residues that need to be evaluated for the respective $\mathcal{M}$-matrix elements.}
\label{tab:zvY}
\end{table}
The cases $S(\mathcal{Y}_{33})$, $S(\mathcal{Y}_{44})$ and $S(\mathcal{Z}_{44})$ are special in the sense that $\tilde{\mu}$ should be split into two separate parts due to the numerator $(u^- - v^+)$. That being said, evaluating the integration over either $u$ or $v$ first results in 
\begin{equation}
S(\mathcal{M})^{u^-}=S(\mathcal{M})^u_{\epsilon} +S(\mathcal{M})^v_{\psi}
\end{equation}
or
\begin{equation}
S(\mathcal{M})^{v^+}=S(\mathcal{M})^v_{\epsilon} +S(\mathcal{M})^u_{\psi}
\end{equation}
The first term on each right hand side represents the contribution where the derivative resulting from the double pole acts on the integrand in \eqref{eq:contributionM} and the residue in the second variable comes from a simple pole of $\tilde{\mu}$. Instead of taking the derivative straight away, we regularise the double pole in, say, $u$ like
\begin{equation}
\int_{\mathcal{C}}  \text{d}u \frac{1}{\left(u-i\frac{K}{2}\right)^{2}} f(u) \rightarrow \int_{\mathcal{C}} \text{d}u \frac{1}{\left(u-i\frac{K}{2}\right)\left(u-i\left(\frac{K}{2}+\epsilon\right)\right)} f(u) 
\end{equation}
where $f(u)$ is without any pole structure and ${\mathcal{C}} $ is a contour containing the pole. We are left with two single poles $u_{0}^{\text{I}}=i\frac{K}{2}$ and $u_{0}^{\text{II}}=i\left(\frac{K}{2}+\epsilon\right)$.  The sum of the relevant residues then takes the simple form
\begin{equation}
\text{res}(f)\big|_{u_{0}^{\text{I}}}+\text{res}(f)\big|_{u_{0}^{\text{II}}} =-\frac{f(i\frac{K}{2})}{\epsilon} + \frac{f(i\frac{K}{2}+i\,\epsilon)}{\epsilon}
\label{eq:PoleEps}
\end{equation}
We simply work up to leading order in $O(\epsilon^0)$ in the second term. Contributions of order $O(\epsilon^{-1})$ vanish due to the sign difference of the two parts of \eqref{eq:PoleEps}.

On the other hand, in the $S(\mathcal{M})^{\psi}$ contribution the derivative from the double pole of the first integration acts on the denominator $\Gamma[1+\tfrac{K+L}{2} + i(u-v)]$ of $\Sigma^{KL}$. For instance, starting with the double pole $1/(u^-)^2$ the phase becomes
\begin{equation}
-i \frac{\Gamma[1+K + L/2 + i \, v]}{\Gamma[1+K] \Gamma[1+L/2 + i \, v]} \, \psi[1 + L/2 - i \, v] \label{dPhase}
\end{equation}
at  $u \eqsp i \, K/2$. We can now pick residues of the digamma function at $v \eqsp - i(L/2 + j), \, j \in \mathbb{N}$. Notice that the residue $j \eqsp 0$ from the measure is excluded here. 

The re-summation of $S(\mathcal{M})^v_{\psi}$ is less than straightforward even for the simplest matrix element $\cX$, cf. \cite{usInter}. In the following two sections we give a detailed discussion of our strategy on the example of the most complicated case.

\section{The hardest computation: $S(\cZ_{33})_\psi^v$}  \label{hardest}
Following the form described in \cite{ETH}, the new form of each $\cZ$ elements is
\beq
\cZ \eqsp G_1 * R_1 . \cX + G_2 * R_2 . \cX
\eeq
where $G_1, \, G_2$ are simple functions of the Zhukowski variables $x^\pm$ akin to the elements of the scattering matrix for fundamental particles \cite{beisertPsu22}. These contain all the coupling dependence and vary under crossing and mirror transforms. On the other hand, $R_1, \, R_2$ are rational coefficient vectors that are not affected by crossing. 

In particular, for any $\cZ_{33}$ element
\beq
G_1 \eqsp \frac{(x_1^+-x_2^+)^2}{U_1 \, U_2 \, (x_1^+ - x_2^-)(x_1^--x_2^+)} \, , \qquad G_2 \eqsp \frac{U_1 \, U_2 \, (x_1^- - x_1^+)(x_2^- -x_2^+)}{(x_1^+ x_2^+)^2 (1 - 1/(x_1^+ x_2^+))^2} \, , \qquad U_i \eqsp \sqrt{\frac{x_i^+}{x_i^-}}
\eeq
and on the diagonal with $n \eqsp k$
\begin{eqnarray}
R_1 & = & \frac{1}{(\dupp)^2} \left\{ - (K-k) \, l, \, - (L-l) (\dupp - k), \, \frac{(K-k) \, l \, (\dupp + L - l)}{L-l}, \, (\dupp - k)(\dupp + L - l) \right\} \, , \notag \\
R_2 & = & \frac{1}{K \, L}  \left\{ -(K-k) \, l, \, -(L-l) \, (-k), \, (K-k) \, l, \, (-k) (L-l) \right\} \,  , \quad \dupp = - i \left(u + \frac{i \, K}{2} - v - \frac{i \, L}{2} \right) \, . \label{R12}
\end{eqnarray}
Combined with the Zhukowski dependent pre-factor $D$ of the $\cX$ matrix
\beq
\{ G_1 *D, \, G_2 * D\} \ \stackrel{3 \gamma, \ 1 \gamma}{\longrightarrow} \, \frac{(u_1^+)^2 \, u_2^- u_2^+}{g^2} \left\{ \frac{1}{(u_1^+-u_2^-)^2}, \, \frac{u_1^- - u_2^+}{(u_1^+ - u_2^-)(u_1^+ - u_2^+)^2} \right\} + \ldots \, . \label{GD}
\eeq
We will close the contour for the $u,v$ integration over the upper and lower half-plane respectively, because in that way $u-v$ can pick up residues only the in the upper half-plane, while the $\cX$ matrix has singularities in $u-v$ only in the lower half-plane; the same is obviously true for the poles in $x_1^+-x_2^-$ in the last formula. However, the double pole in $\dupp$ is obviously in the upper half-plane whenever $K \, < \, L$, so it would seem we have to take into account residues from here. 

Note that $R_2 . \cX$ has an explicit numerator factor $\dupp$ whenever $K \, \leq \, L$. Since the numerator of $R_1$ reduces to that of $R_2$ when $\dupp \eqsp 0$ it follows that $Z_{33}$ has maximally a first order pole in $\dupp$. More is true: in the sum of all terms the $\dupp$ singularity cancels. In fact, in the $3 \gamma, \, 1 \gamma$ kinematics the only $u-v$ singularities of $\cZ_{33}$ are simple poles in the lower half-plane, just like those of $\cX$ itself. It is a shortcoming of the form of the $\cZ$ matrix used here that it does not make these properties manifest. However, we will exploit in the following that the coefficients $G_1, \, G_2, \, R_1, \, R_2$ are simple enough to be absorbed into the $\Gamma$ functions defining the various $\cX$ matrices.

Recall that the measure \eqref{defM} contributes a further factor
\beq
M \eqsp \frac{g^4 \, K \, L}{(u^+ \, u^- \, v^+ \, v^-)^2} + \ldots \nonumber
\eeq
to leading order in $g$. Thus the $\cZ_{33}$ contribution comes in at $O(g^2)$ as expected for a one-loop process. Second, combining the denominator of the last formula with the numerator outside the braces in \eqref{GD} we are left with the singularities
\beq
\tilde \mu \eqsp \frac{1}{(u^-)^2 \, v^- \, v^+} \, . \label{localPoles}
\eeq
as stated in Table 1. We cannot localise both integrations using the numerator $\Ga$ functions of the mirror/mirror BES phase \eqref{sig11}: with $u \eqsp i (m + K/2), \, v \eqsp - i (n + L/2)$ the denominator factor $\Ga[1+(K+L)/2 + i (u-v)]$ is also singular. On the other hand, if we localise $v \eqsp -i L/2$ from the $v^+$ denominator of the last equation, the phase collapses to $\Ga[1+K/2 + i \, u + L]/(\Ga[1+K/2 - i \, u] \Ga[1+b]$ and thus there is no more numerator with a $u$ singularity in the upper half-plane. 

However, starting on the double pole of \eqref{localPoles} at $u \eqsp i \, K/2$ the situation changes: partially integrating $\partial_u$ onto the very denominator $\Ga[1+(K+L)/2 + i (u-v)]$ the digamma function $\psi[1 + L/2 - i \, v] $ appears (see \eqref{dPhase}) whose residues at $v \eqsp - i(L/2 + j), \, j \in \mathbb{N}$ we must pick. The resulting replacement
rule is
\beq
u \rar  i \, \frac{K}{2} \, , \qquad v \rar -i \left(\frac{L}{2} + j \right) \, , \ j \in \mathbb{N}
\eeq
whereby $v^- \rar -i \, (L+j), \, v^+ \rar - i \, j, \, \dupp \rar K+ j$ and \eqref{dPhase} turns into
\beq
- \frac{\Ga[1+K+L+j]}{\Ga[1+K] \Ga[1+L+j]} \, .
\eeq
The resulting top complicated six-parameter series of residues is the topic of this section.

Including the weight factor $W$ (and omitting $g^2$) we obtain
\begin{eqnarray}
W * M * G_1 * D * \Sigma & \rightarrow & \frac{z^{K-k} \, b^k \, y^{L-l+j} \, a^{l+k} \ L \, \Ga[j+K+L]}{j \, (j+L) \, (j+K+L) \, \Ga[K] \, \Ga[1+j+L]} \, , \label{WMGDS} \\
W * M * G_2 * D * \Sigma & \rightarrow & \frac{z^{K-k} \, b^k \, y^{L-l+j} \, a^{l+k} \ L \, \Ga[j+K+L] }{(j+K)^2 \, (j+L) \, \Ga[K] \, \Ga[1+j+L]} \nonumber
\end{eqnarray}
and
\beq
R_1 \rar \frac{1}{(j+K)^2} \left\{ - (K-k) \, l, \, - (L-l) (j+K - k), \, \frac{(K-k) \, l \, (j+K + L - l)}{L-l}, \, (j+K - k)(j+K+ L - l) \right\}
\eeq
whereas $R_2$ remains unchanged.

The summation ranges of the counters are $K \, > \, 1, \, 0 \, < \, k \, < \, K, \, L \, > \, 0, \, 0 \, \leq \, l \, \leq \, L$ and for the internal counter $m$ in $\cX^{kl}_n$ one has $0 \, \leq \, m \, \leq \, k$ (cf. \eqref{defX}). The sum $j \, > \, 0$ is already independent. To decouple the $k, \, l$ sums from $K, \, L$ one will use the shifts $K \rar K+k$, $L \rar L + l$ --- taking out a term $L \eqsp 0 \eqsp l$ --- so that $K, k \, > 0, \, L, l \, \geq \, 0$. Here the case $l \eqsp 0$ (formerly $L \eqsp l$) requires some care as it is singular without a regulator $L \rar L+\ep$. This will create an extra term only for the $\cX^{k,l-1}_{n-1}$ part of $R_1$.

Further, it turns out \cite{usFivePoints} to be useful to map $m \rar k-m$. However, there is a subtlety as our spanning system has $\cX$ matrices with upper index $k-1$ instead of $k$ so that  the map becomes $m \rar k-1-m$. One will finally try to decouple $m$ from $k$ by a replacement of the type $k \rar k+m$. Here we must once again pay attention to the shifted sum range for $m$ where needed: for $\cX$ matrices with index $k-1$ the new range is $k \, > \, 0, \, m \, \geq \, 0$, otherwise we may put $k \rar k-m-1$ and sum over $k \, > \, 0, \, m \, \geq \, 0$ though taking out  the $k \eqsp 1, \, m \eqsp 0$ terms where necessary. In fact, they are non-vanishing solely for the $\cX^{k,l}_n$ part of $R_1$.

We express the Pochhammer symbols in the $\cX$ matrices \eqref{defX} by ratios of $\Ga$ functions and shift the counters as explained replacements. One now encounters the singular ratio $\Gamma[-k - L - m - j]/\Ga[- k - l - L - j]$ --- mostly with shifts of the arguments with small integers --- in every $\cX$ matrix. To make sense of it we momentarily assume $j$ to be non-integer and apply the doubling identity for the $\Ga$ function to transform to positive arguments and a ratio of $\sin$ functions. The latter yields $(-1)^{l+m}$ by the addition theorem.

In the denominator of the expressions so obtained we still find $\Ga[L-m]$ or $\Ga[1+L-m]$ and $\Ga[l-m]$ or $\Ga[1+l-m]$. Since a divergent denominator $\Ga$ functions annihilates the summand, it is in all cases correct to apply a final shift $L \rar L + m, \, l \rar l+m$. Now there is no more \emph{sign mixing} --- all counters occur in the $\Ga$ functions or rational factors from \eqref{WMGDS}, $R_1, R_2$ with positive sign only. 

The eight blocks from $R_1 . \cX, \, R_2 . \cX$ (including \eqref{WMGDS}) are included in the ancillary material to the publication. Those pertaining to $G_1 R_1 . \cX$ have the rational terms\footnote{There may be some further rational numerator terms without $l$.}
\begin{eqnarray}
&& \frac{l + L + 2 \, m}{j \, (j + k + K + m) (j + l + L + 2 \, m) (j + k + l + K + L + 3 \, m)} \notag \\
& = & \frac{1}{k+K+m} \left[\frac{1}{j \, (j + k + l + K + L + 3 \, m)} + \frac{1}{(j + k + K + m) (j + l + L + 2 \, m)} \right]
\end{eqnarray}
with $k \rar k-1$ for the third and fourth block due to the replacements for $X$ matrices with index $k$. We will call the first and second $l$ denominator in the square brackets type A and type B, respectively. Blocks \# $5, \, 6, \, 7, \, 8$ are always only of type B as is recognisable from \eqref{WMGDS}, \eqref{R12}.

Blocks \# 1A and \# 2A can conveniently be combined:
\beq
S_{12A} \eqsp \frac{z^{K} \text{b}^{[km]} a^{[jlm]} y^{[jLm]} \, \Gamma_{[km]} \, \Gamma_{1,lm} \, \Gamma_{1,Km} \,
   \Gamma_{2,Lm}  \, \Gamma_{jkKm} \, \Gamma_{1,jKLm} \, \Gamma_{jklL2m}}{j \, [jklKL3m] \, \Gamma_k \, \Gamma_l \,  \Gamma_K \, \Gamma_{1,L} \, \Gamma_{1,m} \, \Gamma_{2,m} \, \Gamma_{1,jKm} \, \Gamma_{1,kKm} \, \Gamma_{1,jkL2m} \, \Gamma_{1,jlL2m} } \label{S12A}
\eeq
where $\Gamma_{1,jkL2m} \eqsp \Ga[1+j+k+L+2 \, m], \, [km] \eqsp k+m$ etc. 

The counters $j,\, k, \, K, \, l, \, L, \, m$ each come in a \emph{balanced} fashion: for instance, $k$ occurs in three numerator $\Ga$ functions and in three denominator $\Ga$'s with coefficient 1. Let us associate $\underline{x} \eqsp \{1,0,0,0,1,0,1\}$ with the coefficient of $k$ in the argument of the seven numerator $\Ga$ functions, and likewise $\underline{y} \eqsp \{1,0,0,0,0,0,0,1,1,0\}$ with the denominator. Then $\Delta_k \eqsp \sum x_i - \sum y_i \eqsp 0$. Rational factors like $[jklKL3m]$ do not touch upon this counting as they can be written as ratios of $\Ga$ functions. It is then clear that every individual sum in \eqref{S12A} is a hypergeometric $_{p+1}F_{p}$. Note that this property is realised also in sums over residues of Mellin-Barnes representations \cite{MB} of Feynman integrals, cf. \cite{conicHulls} .

The simplest individual sum is the one over $l$ yielding
\beq
(_3F_2)^{\{[2,m], [1,jkL2m], [1,jkKL3m]\}}_{\{[2,jL2m], [2, jkKL3m]\}}[a]
\eeq
multiplied by $\Ga$ functions and rational factors. The $_{p+1}F_p$ functions obey a recursive relation: denote the upper and lower indices by vectors $\underline{a}, \, \underline{b}$. Now select the components $i,\, j$ of these vectors and let $\hat{\underline{a}}, \, \hat{\underline{b}}$ denote the original coefficient vectors with the $i$-th and $j$-th component omitted, respectively. Then
\beq
(_{p+1}F_p)^{\underline{a}}_{\underline{b}}[x] \eqsp \frac{\Ga[b_j]}{\Ga[a_i] \Ga[b_j-a_i]} \int_0^1 \text{d}\mathbf{q} \, (_pF_{p-1})^{\hat{\underline{a}}}_{\hat{\underline{b}}}[\mathbf{q} \, x] \, . \label{reducePFP}
\eeq  
The hypergeometric function is symmetric under the exchange of the upper and the lower labels amongst themselves, respectively. However, when these are not all equal, up to $p \, (p+1)$ inequivalent integral representations are generated. The art is to pick the labels used for the decomposition so that:
\begin{itemize}
\item no denominator $\Ga$ with negative integer argument arises - this corresponds to a divergent integration,
\item the new $\Ga$ functions in \eqref{reducePFP}  cancel the largest possible number of terms in \eqref{S12A}.  
\end{itemize}
In the case at hand this suggests to choose $(i,j) \eqsp (3,2)$ in first step and $(i,j) \eqsp (1,1)$ in a second, or to start on $(1,1)$ and then use $(2,1)$. These two ideal representations are simply related by exchanging the integration parameters $\mathbf{q},\, \mathbf{r}$. The second choice yields
\begin{eqnarray}
S_{12A} & = &\int_0^1 \text{d}\mathbf{q} \, \text{d}\mathbf{r} \, z^K \, b^{[km]} \, y^{[jLm]} \, a^{[1,jm]} \, \mathbf{q}^{[1,m]} \, (1-\mathbf{q})^{[-1,jLm]}  \, \mathbf{r}^{[jkKL3m]} \, (1-a \, \mathbf{q} \, \mathbf{r})^{-[1,jkL2m]} \, * \label{nol} \\ &&\quad \ \ \frac{\Gamma_{km} \, \Gamma_{1,Km} \, \Gamma_{jkKm} \, \Gamma_{2,Lm} \,  \Gamma_{1,jKLm}}{j \, \Gamma_k \, \Gamma_K \, \Gamma_{1,L} \, \Gamma_{1,m} \, \Gamma_{2,m} \, \Gamma _{1,jKm} \,  \,\Gamma_{1,kKm} \, \Gamma_{jLm}} \notag
\end{eqnarray}
Importantly, this process may be iterated \cite{usFivePoints} because the elements of the last formula are of the same type as before, just that polynomial or rational arguments are introduced. We will thus successively build up more and more complicated rational functions under a multiple integral while reducing the number of $\Ga$ functions and rational factors.

Taylor expanding $(1- a \, \mathbf{q} \, \mathbf{r})^{-[1jkL2m]}$ under the integral and taking the integrations gives back  the original series \eqref{S12A}; here this simply proves the equivalence of the integral form of $_3F_2$ and its series representation. In successive steps this is a non-trivial test because we are swapping integrations and sums. Nevertheless, for the finite integrals we are dealing with this test always works. This provides a very valuable guideline. 

None of the remaining sums in \eqref{nol} is of geometric type. It seems logical to choose $k$ or $L$ for the second sum to take because they are the next simplest; we will opt for $L$. Let $s$ the integration parameter of the resulting $_2F_1$. A new subtlety arises: we find
\beq
(_2F_1)^{\{[2,m], [1,jKm]\}}_{\{[jm]\}}\left[\frac{(1-\mathbf{q}) \, \mathbf{r} \, y}{1 - a \, \mathbf{q} \, \mathbf{r}} \right] 
\eqsp \frac{\Gamma[j+m]}{\Gamma[j-2] \, \Gamma[m+2]} \int_0^1 \text{d}\mathbf{s} \,  \frac{(1-\mathbf{s})^{j-3} \mathbf{s}^{m+1}  (1- a \, \mathbf{q} \, \mathbf{r})^{[1jKm]} }{(1 -a \, \mathbf{q}
   \mathbf{r}+\mathbf{q} \, \mathbf{r} \, \mathbf{s} \, y-\mathbf{r} \, \mathbf{s} \, y)^{[1jKm]}} \label{j1Terms} 
\eeq
where we have factorised powers of the composite argument of the hypergeometric function, which can create problems with cuts. In this article we do not pay much attention to cuts; if necessary they can be corrected at the very end of the computations comparing the Taylor expansion of the results in terms of dilogarithm to the original series of residues.

The second row of the last equation results from decomposing over the labels (2,1) in order to avoid a denominator factor $\Ga[-1-K]$ rendering the $K$ sum problematic in a later step. However, the last formula hides a problem of the very same nature: our next step will in fact be the sum over $j$. It cannot naively be taken from $j \eqsp 1,2,\ldots$ because the first two terms \eqref{j1Terms} come with the divergent integrals $(1-\mathbf{s})^{-2}, (1-\mathbf{s})^{-1}$ which go with the denominator singularities $\Gamma[-1], \, \Ga[0]$. 

In fact, for $j \eqsp 2$, the $L$ sum in \eqref{nol} is manifestly geometric whence no integration parameter will arise. Second, putting $j \eqsp 1$ we obtain
\beq
\sum_{L \eqsp 0}^\infty \frac{\Ga_{2,KLm} \, \Ga_{2,Lm}}{\Ga_{1,L} \, \Ga_{1,Lm}} \left(\frac{(1-\mathbf{q}) \, \mathbf{r} \, y}{1 - a \, \mathbf{q} \, \mathbf{r}}\right)^L
\eeq
which is also only formally a $_2F_1$:
\beq
\frac{\Ga[2 + L + m] \, \Ga[2 + K + L + m]}{\Ga[1+L+m]} \eqsp \Ga[3+K+L+m] - (1+K) \, \Ga[2+K+L+m] \label{reduceJ}
\eeq 
so that two geometric sums are obtained. We keep the cases $j \eqsp1,2$ separate not introducing any integration parameter $s$  there, and let the $j$ sum in \eqref{j1Terms} start at $j \eqsp 3$. It is once again possible to check by Taylor expansion under the integrals that the sum of these three parts equals \eqref{S12A}.

Whenever there is a ratio $\Ga[m+n]/\Ga[n], \, m \in \mathbb{N}$, the reduction of an apparent $_{p+1}F_p$ to a $_pF_{p-1}$ given by a sum over $n$  is generically necessary; this will involve more terms for $m \, > \, 1$. Note that the explicit factor of $(1+K)$ in the second term of the rhs. of \eqref{reduceJ} will eventually be needed in the $k$ sum for both terms to have the same polynomial order. Once introduced, such factors will occur in any later sum.

At $j \eqsp 1,2$ the remaining sums are all geometric; these parts generate two-parameter Euler integrals. We put these aside for now bearing in mind that they do obviously contribute to the final result for $S_{12A}$. Instead we turn to the $j \, > \, 2$ sum. We decompose
\beq
\frac{1}{j \, \Ga[-2+j]} \rar \frac{1}{\Ga[-1+j]} -\frac{2}{\Ga[j]} + \frac{2}{\Ga[1+j]}
\eeq
and let the sum start at 0 which necessitates subtracting that term out of the last part again. All three sums are geometric, the subtraction term is obviously rational. Next, the $k$ sum yields a $_2F_1$ in some of the terms. We decompose over the labels (1,1) creating a fourth integration parameter $t$ there; the other cases remain three-parameter integrals. Finally, the $K$ and $m$ sums are all geometric.

Collecting terms, beyond the $j \eqsp 1,2$ parts $S_{12A}$ is given by four fairly simple Euler-integrals: two of these have all four integration parameters $\mathbf{q}, \, \mathbf{r}, \, \mathbf{s}, \, \mathbf{t}$ and the two others only the first three, so $\mathbf{q}, \, \mathbf{r}, \, \mathbf{s}$. Interestingly the denominators of both classes essentially reduce to powers of the two polynomials
\begin{eqnarray}
\label{eq:P1}
P_1 & = & 1 - b - 2 \, a \, \mathbf{q} + a \, b \, \mathbf{q} + a^2 \, \mathbf{q}^2 - 2 \, a_0 \, y + a_0 \, b \, y + 2 \, a_0 \, \mathbf{q} \, y + 
 2 \, a_0 \, a \, \mathbf{q} \, y - a_0 \, b \, \mathbf{q} \, y - 2 \, a_0 \, a \, \mathbf{q}^2 \, y - 2 \, \mathbf{s} \, y \notag \\
 &+ & 2 \, a_0 \, \mathbf{s} \, y + 2 \, b \, \mathbf{s} \, y - a_0 \, b \, \mathbf{s} \, y + 2 \, \mathbf{q} \, \mathbf{s} \, y - 2 \, a_0 \, \mathbf{q} \, \mathbf{s} \, y + 2 \, a \, \mathbf{q} \, \mathbf{s} \, y - 
 2 \, a_0 \, a \, \mathbf{q} \, \mathbf{s} \, y - 2 \, b \, \mathbf{q} \, \mathbf{s} \, y + a_0 \, b \, \mathbf{q} \, \mathbf{s} \, y - a \, b \, \mathbf{q} \, \mathbf{s} \, y  \notag \\
 & - & 2 \, a \, \mathbf{q}^2 \, \mathbf{s} \, y + 2 \, a_0 \, a \, \mathbf{q}^2 \, \mathbf{s} \, y + a \, b \, \mathbf{q}^2 \, \mathbf{s} \, y + a_0^2 \, y^2 - 
 2 \, a_0^2 \, \mathbf{q} \, y^2 + a_0^2 \, \mathbf{q}^2 \, y^2 + 2 \, a_0 \, \mathbf{s} \, y^2 - 2 \, a_0^2 \, \mathbf{s} \, y^2 - a_0 \, b \, \mathbf{s} \, y^2  \notag \\
& - &4 \, a_0 \, \mathbf{q} \, \mathbf{s} \, y^2 + 4 \, a_0^2 \, \mathbf{q} \, \mathbf{s} \, y^2 + 2 \, a_0 \, b \, \mathbf{q} \, \mathbf{s} \, y^2 + 2 \, a_0 \, \mathbf{q}^2 \, \mathbf{s} \, y^2 - 2 \, a_0^2 \, \mathbf{q}^2 \, \mathbf{s} \, y^2- a_0 \, b \, \mathbf{q}^2 \, \mathbf{s} \, y^2 + \mathbf{s}^2 \, y^2 - 2 \, a_0 \, \mathbf{s}^2 \, y^2 \notag \\
& + & a_0^2 \, \mathbf{s}^2 \, y^2 - b \, \mathbf{s}^2 \, y^2 + a_0 \, b \, \mathbf{s}^2 \, y^2 - 2 \, \mathbf{q} \, \mathbf{s}^2 \, y^2 + 4 \, a_0 \, \mathbf{q} \, \mathbf{s}^2 \, y^2 - 2 \, a_0^2 \, \mathbf{q} \, \mathbf{s}^2 \, y^2 + 2 \, b \, \mathbf{q} \, \mathbf{s}^2 \, y^2 - 2 \, a_0 \, b \, \mathbf{q} \, \mathbf{s}^2 \, y^2 \label{defP1} \\
& + & \mathbf{q}^2 \, \mathbf{s}^2 \, y^2 - 2 \, a_0 \, \mathbf{q}^2 \, \mathbf{s}^2 \, y^2 + a_0^2 \, \mathbf{q}^2 \, \mathbf{s}^2 \, y^2 - b \, \mathbf{q}^2 \, \mathbf{s}^2 \, y^2 + a_0 \, b \, \mathbf{q}^2 \, \mathbf{s}^2 \, y^2 - z + b \, z + 2 \, a \, \mathbf{q} \, z - a \, b \, \mathbf{q} \, z - a^2 \, \mathbf{q}^2 \, z \notag \\
& + & a_0 \, y \, z - a_0 \, \mathbf{q} \, y \, z - a_0 \, a \, \mathbf{q} \, y \, z + a_0 \, a \, \mathbf{q}^2 \, y \, z + \mathbf{s} \, y \, z - a_0 \, \mathbf{s} \, y \, z - b \, \mathbf{s} \, y \, z - \mathbf{q} \, \mathbf{s} \, y \, z + a_0 \, \mathbf{q} \, \mathbf{s} \, y \, z - a \, \mathbf{q} \, \mathbf{s} \, y \, z \notag \\
& + & a_0 \, a \, \mathbf{q} \, \mathbf{s} \, y \, z + b \, \mathbf{q} \, \mathbf{s} \, y \, z + a \, \mathbf{q}^2 \, \mathbf{s} \, y \, z - a_0 \, a \, \mathbf{q}^2 \, \mathbf{s} \, y \, z \, , \notag \\[2 mm]
P_2 & = & 1 - a \, \mathbf{q} - a_0 \, y + a_0 \, \mathbf{q} \, y - \mathbf{s}  \, y + a_0 \, \mathbf{s} \, y + \mathbf{q} \, \mathbf{s} \, y - a_0 \, \mathbf{q} \, \mathbf{s} \, y - z + a \, \mathbf{q} \, z \, .
\end{eqnarray}
Note that $P_1$ is linear in $b, \, z$ and quadratic in all other variables while $P_2$ is even multilinear. In detail, upon rescaling as
\beq
a \rar a/ \mathbf{r} \, , \quad b \rar b / (\mathbf{r} \, \mathbf{t})\, , \quad y \rar y / \mathbf{r} \, , \quad z \rar z / (\mathbf{r} \, \mathbf{t}) \label{scale4A}
\eeq
followed by $\mathbf{r} \rar a/a_0$, the four-parameter integrand of $S_{12A}$ becomes
\beq
\frac{- 2 \, a_0 \, b \, \fq \, \fs \, z}{(1-\fq) (1-\fs)^3 \, t \, (1 - a \, \fq - \fs \, y + \fq \, \fs \, y - z + a \, \fq \, z) (1 - b - a \, \fq - \fs \, y + b \, \fs \, y + \fq \, \fs \, y - b \,  \fq \, \fs \, y - z + b \, z + a \, \fq \, z)} \notag
\eeq
\beq
+ \frac{2 \, a_0 \, b \, \fq \, \fs \, z \, (1 - a \, \fq - a_0 \, y + a_0 \, \fq \, y - \fs \, y + a_0 \, \fs \, y + \fq \, \fs \, y - a_0 \, \fq \, \fs \, y)}{(1 - \fq) (1 - \fs)^3 \, \ft \, P_2 \, P_1} \, . \label{fourPS12A}
\eeq
For the three-parameter part we scale as
\beq
a \rar a / \mathbf{r} \, , \quad b \rar b / \mathbf{r} \, , \quad y \rar y / \mathbf{r} \, , \quad z \rar z / \mathbf{r} \label{scale3A}
\eeq
also followed by $\mathbf{r} \rar a/a_0$ to find the integrand
\beq
\frac{-a_0^2 \, b \, \fq \, \fs \, y \, z  \, (2 - 2 \, a \, \fq - 3 \, a_0 \, y + 3 \, a_0 \, \fq \, y - 2 \, \fs \, y + 3 \, a_0 \, \fs \, y + 2 \, \fq \, \fs \, y - 
 3 \, a_0 \, \fq \, \fs \, y - 2 \, z + 2 \, a \, \fq \, z)}{(1-\fs)^2 \, P_2^2 \, P_1} +
\eeq
\beq
\frac{a_0^3 \, b \, \fq \, \fs \, y^2 \, z \,  (2 - b - 2 \, a \, \fq - 2 \, a_0 \, y + 2 \, a_0 \, \fq \, y - 2 \, \fs \, y + 2 \, a_0 \, \fs \, y + b \, \fs \, y + 2 \, \fq \, \fs \, y - 2 \, a_0 \, \fq \, \fs \, y - b \, \fq \, \fs \, y - z + a \, \fq \, z)}{(1-\fq)^{-1} \, (1-\fs) \, P_2 \, P_1^2} \, . \notag
\eeq
Undoing the scaling, the sum of these four integrals is finite and can be tested against the original series by Taylor expansion under integral (including the $j \eqsp 1,2$ parts). To this end $a, \, b, \, y, \, z$ are regarded as small wrt. to some common expansion scale. The $\fs$ integral becomes well-defined after the other integrations have been taken and hence ought to be done last. 

The A part of blocks \# 3 and \# 4 can be dealt with in the same way; in both cases a $j \eqsp 1$ term needs to be kept separate; these are of the same type as the $j \eqsp 2$ case above. The result for the $j \, > \, 1$ sum is concise for block \# 3, but not for block \# 4 so that we refrain from showing it here. 

The B parts of all blocks are also individually re-summed into Euler integrals. Again, $j \eqsp 1$ terms always have to be kept separate. The analysis of the $j \, > \, 1$ sum then proceeds much as before; also here in the final sum --- for us the one over $m$ --- irreducible three- and four-parameter polynomials arise that are related to $P_1$ by variable re-scalings. All $\Ga$ functions will have disappeared from the coefficients of the final sums. We will find
\beq
\sum_{m \eqsp 0}^\infty R^m \eqsp \frac{1}{1-R} \, , \qquad \sum_{m \eqsp 0}^\infty m \, R^m \eqsp \frac{R}{(1-R)^2} \, ,  \qquad \sum_{m \eqsp 0}^\infty m^2 \, R^m \eqsp \frac{R \, (1+R)}{(1-R^3)} \, . \label{mSums}
\eeq
Thus we should push down the order of $m$ as a coefficient of the last sum in order to limit to $P_1^{-2}$ the irreducible polynomial $P_1 \sim (1-R)$ to facilitate the intended analysis of the resulting Euler integrals by intersection theory. To this end, we may employ the relation
\beq
\frac{\Ga[j+k+K+m]}{j+k+K+m} \eqsp \Ga[-1+j+k+K+m] \left[ 1 -\frac{1}{j+k+K+m} \right] \label{shiftGamma}
\eeq
(or a shift of the argument by -1, respectively) after the $l, \, L$ summations. As a result, $R$ from \eqref{mSums} --- which is normally a ratio of several polynomials --- will create higher powers of simpler polynomials in the denominator, while $P_1$ only comes to second order. Hence \eqref{shiftGamma} acts in essence as a parametric IBP identity.

Every B part yields an Euler integral with a three-parameter version of $P_1$ and a second one with a four-parameter $P_1$. In the sum over all B parts the four-parameter version cancels. The simple rescaling
\beq
a \rar a/\mathbf{r} \, , \qquad y \rar y/\mathbf{r}
\eeq
followed by $\mathbf{r} \rar a/a_0$ sends the three-parameter polynomial to $P_1$.

Beyond the aforementioned leading $j$ cases there are two more sources of extra terms:
\begin{itemize}
\item Having taken the $l \, \geq \, 0$ sum in both, the A and B parts of block \# 3 there is a boundary contribution from the ratio $\Ga[L+m]/\Ga[m]$ that our Taylor expansion test requires. To pick it up we put $L \rar 0$ first and then $m \rar 0$. Block \# 4 features a similar ratio, namely $\Ga[L+m]/\Ga[L]$ which is lost in this limit, and so is not required to make the test work. In fact, both extra terms evaluate\footnote{taking the opposite limit $m \rar 0$ first and then $L \rar 0$ for block \# 4} to the same sum of two-parameter Euler integrals. The contribution is apparently needed only once to match the $L, \, l \rar 0$ part of $\cZ_{33}$ obtained in the notebook \cite{ETH} from the regulator $L \rar L + \ep$.
\item Block \# 4 has a $k \eqsp 1, \, m \eqsp 0$ part that needs to be taken out in our way of decoupling of the sums. This yields two-parameter integrals, too. 
\item Note that the $j \eqsp 1,2$ terms from the A sector can be made into two-parameter integrals. However, the B sector $j \eqsp 1$ terms yield three-parameter integrals.
\item All these parts are simple: they do not involve $P_1$.
\end{itemize}

 Eg. our expressions \eqref{scale3A}, \eqref{scale4A} have poles in $1-\fs$, the parameter from the $j$ sums. These are the only divergences of the integrands. To proceed towards an evaluation we partially fraction the sum of all un-scaled integrals in $\fs$. We obtain:
\begin{itemize}
\item two integrals with $P_1$ as a single or double pole, respectively, from the unscaled three-parameter B cases,
\item two integrals with $P_1$ as a single or double pole, respectively, from the unscaled three-parameter A cases,
\item two integrals with $P_1$ as a single or double pole, respectively, from the unscaled four-parameter A cases.
\item After scaling, their denominators contain only the polynomials
\beq
P_3 \eqsp 1 - a \, \fq - y + \fq \, y \, , \qquad P_4 \eqsp 1 - b - a \, \fq - y + b \, y + \fq \, y - b \, \fq \, y - z + b \, z + a \, \fq \, z \label{defP34}
\eeq
and $P_1$. All six integrals are manifestly finite. Upon expanding the last four have fairly large numerators (948 \ldots 2496 summands). The four-parameter A case with a single power of $P_1$ in the denominator is the hardest integral as it has the longest numerator and features a third order pole in $P_4$. In the next section we will comment on an intersection theory computation along the lines of \cite{usInter} that allows us to integrate out $\fq, \, \fs$ yielding rational terms, logarithms and dilogarithms depending on the five variables $\{a_0, \, a, \, b, \, y,  \, z\}$. We scale back the results to obtain functions depending on $\{a, \, b, \, y,  \, z\}$ but also the remaining integration parameters $\fr$ and $\ft$.
\item All other terms are trivial in that their denominator does not contain $P_1$ and can therefore quite straightforwardly be integrated. There are a few terms with poles in $1-\fs$. Since the singularity is linear, the rest of these rational functions does not depend on $\mathbf{s}$. A third order pole of the four-parameter part cancel upon factorisation. Integrating $\ft$ out of its second order pole we straightforwardly achieve cancellation against the second order pole of the three-parameter part (so it is not necessary to integrate out $\fq, \, \fr$ to see this). Similarly, to cancel the first order poles we integrate the four-parameter residue in $\ft, \, \fq$ and the three-parameter one in $\fq$; hence cancellation ensues before integrating out $\fr$.
\item We integrate the well-defined trivial three- and four parameter cases over $\fq, \, \fs$; this is elementary.
\item The intersection theory results for the integrals with $P_1$ denominators are imported. The variables $\fq$ and $\fs$ --- so the first integration parameter of the $_3F_2$ functions from the $l$ sums and the parameter from the $j$ sums --- have already been removed, see below. We undo the scaling so that $\fr$ and $\fr, \, \ft$ reappear, respectively.
\item We now collect all former four-parameter integrals, reduce their sum to a basis of logarithms and dilogarithms (and a rational part) and integrate over $\ft$. Next, the result combined with all former three-parameter cases is written wrt. a basis of (di)logarithmic functions (and a rational part) then integrated over $\fr$. We could have met quadrilogs, but due to higher order denominators the logarithm weight remains limited to two; higher functions than dilogarithms are never generated.
\item In the above we have omitted the trivial two-parameter part, ie. A type small $j$ terms, the $L \eqsp 0 \eqsp l$ and the $k \eqsp 1, \, m \eqsp 0$ part. Obviously, the integral of the two-parameter part cannot contain higher functions than dilogarithms. The B type small $j$ terms were also left aside, because their third Euler parameter has a different origin as it comes from the $k$ sum. Due to higher order denominators these also remain limited to dilogarithm level.
\item Rational parts and single logarithms cancel upon adding these integrals to the previous sum. The symbol of the dilogarithm function for $S(\cZ_{33})_\psi^v$ is given below.
\item Prior to scaling, the relevant irreducible denominator of B type is quadratic in the three integration parameters $\fq, \,\fr, \, \fs$, the $A$ type denominators are even cubic in $\fr$, and maximally quadratic in $\fq, \, \fs$ or $\fq, \, \fs, \, \ft$, respectively. The dependence on the outer variables $b, \, z$ is linear and that on $a, \, y$ quadratic; after scaling also that on $a_0$ is quadratic. The intersection theory methods of the next section would allow us to handle quadratic dependence on all integration parameters; for the cubic dependence on $\fr$ we would have to enlarge our toolbox, though. This is one reason for the mixed approach of scaling out $\fr, \, \ft$ and running the intersection theory on $\fq, \, \fs$ exclusively. The second argument here is the size of the bases: for the problem outlined here the bases have length 13; if we want to deal with all $\cY$ and $\cZ$ residue sums in a unified way a further denominator polynomial \eqref{defP9} appears that drives up the basis length to 16 for the bivariate intersection theory problem. Renouncing on scaling and trying to run intersection theory on all four parameters the length of the full bases will even increase to 26.

We will derive five canonical Pfaffian equations for the variables $\{a_0, \, a, \, b, \, y, \, z\}$ and solve these up to dilogarithm level. Yet, the relevant equations have denominators quadratic in $b, \, y, \, z $  which could obviously cause square root arguments upon integration. Due to the simplicity of the vector of starting values \eqref{lowestGamma} the problem does not bite in the first iteration step, but it could stop us from attempting the second iteration without variable changes. Fortunately the issue can sidestepped: all quadratic dependence arises in letters like $D_5$ in \eqref{fiveDens} below. If we choose the $a_0$ or $a$ equation as the first to be integrated, both, in the first and second iteration of the set of equations, then all quadratic dependence is already contained in the result of that integration. The other differential equations will thus linearise.

Similarly, upon scaling back, all denominators and arguments of logarithms or polylogarithms will be linear in $\ft$ first (where appropriate) and having taken that integration (where appropriate) also in $\fr$. 

\item It is possible to separately re-sum the A type series of residues for blocks \#1, \#2. Yet, upon doing so linearity of the $\ft, \, \fr$  integrations will be lost in that sector. Combining the two blocks avoids this technical obstacle which makes our analytical evaluation work smoothly. The observation is somewhat aleatory --- it would be better to have a starting point from which this technical problem never arises. In Appendix A we report on a second attempt based on yet another writing of the $\cZ_{33}$ element. The computation has definite advantages but now is harder to see in which sequence to take the sums. For example,  it is not too easy to construct a small Euler representation for the $k \eqsp 1, \, m \eqsp 0$ terms. In one way of summing there are maximally three parameters, but the integration creates trilogarithms at an intermediate stage. These cancel according to the weight three symbol as they must by uniform transcendentality \cite{lipKot} and the end result is identical to the one presented below. We did eventually succeed in finding a two-parameter Euler representation thereby avoiding the issue.
\end{itemize}

Let
\begin{eqnarray}
D_1 & = & a - y - a \, z + b \, y \, , \notag \\
D_2 & = & b - z + a \, z - b \, y \, , \notag \\
D_3 & = & a \, b - y \, z - a \, b \, z + b \, y \, z \, , \\ \label{fiveDens}
D_4 & = & a \, b - y \, z - a \, b \, y - a \, b \, z + a \, y \, z + b \, y \, z \, , \notag \\
D_5 & = & a \, b - y \, z - 2 \, a \, b \, y - a \, b \, z  + 2 \, a \, y \, z + b \, y \, z + a \, b \, y^2  - a^2 \, y \, z \, . \notag
\end{eqnarray}
The symbol of the digamma series of residues for $Z_{33}$ is
\beq
{\cS}_{12A} \eqsp \left\{ 1, \, \frac{b \, (1 - y)}{D_2}, \, \frac{a \, b \, (1 - z)}{D_3}, \frac{a \, b \, (1 - y - z)}{D_4} \right\} \, . \, \{ {\cS}_1, \, {\cS}_2, \, {\cS}_3, \, {\cS}_4 \} \label{prodZ33Sym}
\eeq
where
\begin{eqnarray}
 {\cS}_1 & = & {\cS}\left[1-a,\frac{(1-z) (1-y-z)}{(1-a \, y-z)^2}\right] + {\cS}\left[1-b,\frac{(1-b) (1-y)}{1-a \, y - b}\right] + {\cS}\left[1-a-b,\frac{(1-a \, y-b)^2}{(1-b)^2 (1-y)}\right] + \notag \\ &+& {\cS}\left[y,\frac{(1-z) (1-a \, y-b)}{(1-b) (1-a \, y-z)}\right] + {\cS}\left[b-z,\frac{(1-z) (1-a \, y-b)}{(1-b) (1-a \,y-z)}\right] + {\cS}\left[D_4,\frac{(1-a) (1-b) (1-y-z)}{(1-y) (1-z)
   (1-a-b)}\right] + \notag \\ & + & {\cS}\left[D_5,\frac{(1-y) (1-a-b) (1-a \, y-z)}{(1-a)(1-y-z) (1-a \, y-b}\right] \, , \notag \\
{\cS}_2 & = & {\cS}\left[1-a,\frac{1-a \, y-z}{1-z}\right] + {\cS}\left[1-y,\frac{1-b}{1-a \,y-b}\right] + {\cS}\left[1-a-b,\frac{1-b}{1-a \, y-b}\right]  \label{symS1S2} \\ & + & {\cS}\left[1-y-z,\frac{1-a \, y-z}{1-z}\right]  + {\cS}\left[b-z,\frac{(1-b) (1-a \, y-z)}{(1-z) (1-a \,y-b)}\right] + {\cS}\left[D_4,\frac{(1-y) (1-z) (1-a-b)}{(1-a) (1-b)(1-y-z)}\right] \notag \\ &+& {\cS} \left[D_5,\frac{(1-a) (1-y-z) (1-a \, y-b)}{(1-y)(1-a-b) (1-a \, y-z)}\right] \, , \notag \\
{\cS}_3 & = & {\cS}\left[1-b,\frac{1-a-b}{(1-b)(1-y)}\right] + {\cS}\left[1-z,\frac{(1-a)
   (1-z)}{1-y-z}\right] + {\cS}\left[a,\frac{(1-a)(1-b)}{1-a-b}\right] \notag \\
   & + & {\cS}\left[y,\frac{1-y-z}{(1-y) (1-z)}\right] + {\cS}\left[b,\frac{(1-a)(1-b)}{1-a-b}\right]  + {\cS}\left[z,\frac{1-y-z}{(1-y)(1-z)}\right] \notag \\ &+& {\cS}\left[a-y,\frac{1-a}{1-y}\right] + {\cS}\left[D_1,\frac{(1-b) (1-y-z)}{(1-z)(1-a-b)}\right] +{\cS}\left[D_4,\frac{(1-y) (1-z) (1-a-b)}{(1-a) (1-b)(1-y-z)}\right] \, , \notag \\
{\cS}_4 & = & {\cS}\left[1-a,\frac{1-a \, y-z}{1-y-z}\right] + {\cS} \left[1-b,\frac{1-a \, y-b}{1-a-b}\right] + {\cS}\left[1-a-b,\frac{(1-b) (1-y)}{1-a \, y-b}\right] \notag \\ &+& {\cS}\left[1-y,\frac{1-a-b}{1-a \, y-b}\right] + {\cS}\left[1-z,\frac{1-y-z}{1-a \, y-z}\right] +{\cS}\left[1-y-z,\frac{1-a \, y-z}{(1-a)(1-z)}\right]  \label{symS3S4} \notag \\ & + & {\cS}\left[a,\frac{(1-a-b) (1-a \, y-z)}{(1-a) (1-z) (1-a \, y-b)}\right] + {\cS}\left[b,\frac{1-a-b}{(1-a)(1-b)}\right] + {\cS}\left[a-y,\frac{1-y}{1-a}\right] \notag \\ & + & {\cS}\left[y,\frac{(1-b)
   (1-y) (1-a \, y-z)}{(1-y-z) (1-a \, y-b)}\right] + {\cS}\left[z,\frac{(1-y)(1-z)}{1-y-z}\right] + {\cS}\left[b-z,\frac{(1-b) (1-a \,y-z)}{(1-z) (1-a \, y-b)}\right] \notag \\ & + & {\cS}\left[D_1,\frac{(1-z) (1-a-b)}{(1-b)(1-y-z)}\right]+ {\cS}\left[D_5,\frac{(1-a) (1-y-z) (1-a \,  y-b+1)}{(1-y)(1-a-b) (1-a \, y-z)}\right] \notag
\end{eqnarray}
We remark that the symbols ${\cS}_2, {\cS}_3, {\cS}_4$ change sign under conjugation, likewise $D_2, D_3, D_4$. On the other hand, conjugation changes the numerator of each rational factor in \eqref{prodZ33Sym} into the other half of the associated denominator. Correspondingly, $({\cS}_1)^* - {\cS}_2 - {\cS}_3 - {\cS}_4 \eqsp {\cS}_1$ so that the entire symbol is real. Corresponding dilogarithmic functions are provided in the ancillary material to this publication. Note that the special structure of the rational coefficients in \eqref{prodZ33Sym} --- ie. the numerator terms are summands of the denominator ---  was taken into account in the ansatz for fitting in \cite{usFivePoints} in order to guarantee linear independence.

\section{Elements of bivariate intersection theory} \label{bivariate}

Consider rational 2-forms
\beq
\phi_L \eqsp \hat \phi_L(\fq,\fs) \, \text{d}\fs \wedge \text{d}\fq \, , \qquad \phi_R \eqsp \hat \phi_R(\fq,\fs) \, \text{d}\fs \wedge \text{d}\fq \, ,
\eeq
whose denominator is built out of non-negative powers of 
\beq
\{\fq, \, 1-\fq, \, \fs, \, 1-\fs, \, P_1, \, P_3, \, P_4 \}
\eeq
where the last three polynomials are defined in \eqref{defP1}, \eqref{defP34}. These forms are also allowed to have an arbitrary polynomial numerator depending on $\{a_0, \, a, \, b, \, y, \, z, \, \fq, \, \fs\}$. The aim is to compute integrals in the family
\beq
I \eqsp \iint_0^1 u(\fq,\fs) \, \phi_L \, .
\eeq
The factor $u$ defining the \emph{twisted cohomology} is the product of all possible denominators raised to some small non-integer power $\gamma$:
\beq
u \eqsp \fq^\gamma \, (1-\fq)^\gamma \, \fs^\gamma \, (1-\fs)^\gamma \, P_3^\gamma \, P_4^\gamma \, P_1^\gamma 
\label{eq:u13}
\eeq
For our purposes, $\gamma$ will be the same for all factors of $u$ and simply play the r\^ole of a dimensional regulator. 

Intersection theory \cite{inter1} allows us to decompose the most general integral in the family in terms of a set of masters. Let
\beq
\hat \omega_\fs \eqsp \partial_\fs \, \log(u) \, , \qquad \hat \omega_\fq \eqsp \partial_\fq \, \log(u) \, . \label{countMs}
\eeq
The number of solutions to the simultaneous conditions $\hat \omega_\fs \eqsp 0 \eqsp \hat \omega_\fq$ counts the master integrals \cite{howManyMasters}.

We will employ the \emph{fibration algorithm} of \cite{inter2} in which the integrations are successively done. Let us pick one of the integrations to start with, say, in the variable s which occurs in fewer of the candidate denominator polynomials than $\fq$. Replacing any form $\phi_L$ as some coefficients only depending on $\fq$ times basis one-forms in $\fs$, we should be able to express its integral in terms of a basis of masters. Such a relation does in general not hold at the integrand level.

The relevant count of masters is given by the number of zeroes of the first condition in $\eqref{countMs}$ only. In our application there will be three masters $\{e_1, \, e_2, \, e_3\}$. If anything works, it must be so that
\beq
\phi_L \eqsp \sum_{i \eqsp 1}^3 \left(\phi_L'(\fq)_i \, \text{d}\fq \right)  \wedge e_i  
\eeq
upon integration and similarly 
\beq
\phi_R \eqsp \sum_{j \eqsp 1}^3 h_j \wedge \left(\phi_R'(\fq)_j \, \text{d}\fq \right)
\eeq
for three right masters $h_i$. In our application it will be most convenient to identify the left and right bases. We choose the forms
\beq
\left\{ \frac{1}{\fq \, \fs}, \, \frac{1}{\fq \, (1-\fs)}, \, \frac{1-\fq}{P_1} \right\} \label{basS}
\eeq
The factors $\fq, \, 1-\fq$ are not necessary at this stage; they will rather ensure homogeneous scaling at infinity when $\fq$ is integrated out later on. 

The \emph{intersection number} (in our case a $3 \times 3$ matrix) is a pairing
\beq
C_\fs \eqsp \la e_i | h_j \ra \, .
\eeq
To compute it we determine the total set of poles $\{\fs_1 \ldots \fs_k\}$ in $s$ of the collection of denominators. In our case these are $\{0, \, 1, \, \fs^-, \, \fs^+, \, \infty\}$, where $\fs^\mp$ are the two zeroes of $P_1$ wrt. $\fs$. It is useful to keep these abstract as long as possible and substitute only in the end. Infinity is included as a point to expand around because the problem is of projective nature.

For each pole we replace $\fs \rar \fs_m + \fv$ with a new variable $\fv$ for Taylor expanding. The covariant derivative wrt. to $\fs$ is defined by the partial derivative acting under the integral, where differentiating $u$ causes a connection term:
\beq
\nabla_\fs \eqsp \partial_\fv + \hat \omega_\fs \label{nablaS}
\eeq
We now define a \emph{potential} for every $e_i$ in a neighborhood of a pole $\fs_m$: let us write $e_i \eqsp \hat e_i \, d\fs$, so as in the beginning of the section, hatting expresses the omission of the differential. Assuming the expansion of $\hat e_i|_{\fs \eqsp \fs_m + \fv}$ to start at a minimal power $min_m-1$ and that we can truncate at some maximum $max_m$ we define the potential as
\beq
\psi_{i,m} \eqsp \sum_{n \eqsp min_m}^{max_m} c_i^{(n)} \, \fv^n 
\eeq
and solve
\beq
\left(\partial_\fs + \hat \omega_s|_{\fs \eqsp \fs_m + \fv}\right) \, \psi_{i,m} \eqsp \hat e_i|_{\fs \eqsp \fs_m + \fv} \label{defPsiV}
\eeq
comparing the coefficients of the Taylor expansion of the two sides from $min_m$ to $max_m-1$. The intersection number is now
\beq
C_\fs \eqsp \sum_{m \eqsp 1}^k res|_{\fs \eqsp \fs_m} \, \psi_{i,m} \, \hat h_j|_{\fs \eqsp \fs_m + \fv}
\eeq
It follows that $max_m+1$ is the negative of the minimal order of $\hat h$ in the region $m$. It is advisable to choose the masters such that the range of powers in the potentials is minimised. In practice this implies to prefer masters with  simple poles only, cf. \eqref{basS}. 

Let us expand $P_1 \eqsp c_0 + c_1 \, \mathbf{s} + c_2 \, \mathbf{s}^2$. The discriminant $\Delta_s \eqsp c_1^2- 4 \, c_0 \, c_2$ takes a factorised form:
\beq
\Delta_{\mathbf{s}} \eqsp (1-\fq)^2 \, y^2 \, Q_1
\eeq
where
\begin{eqnarray}
Q_1 & = & a_0^2 \, b^2 - 2 \, a_0 \, a \, b^2 \, \fq + a^2 \, b^2 \, \fq^2 - 2 \, a_0^2 \, b^2 \, y + 
 2 \, a_0^2 \, b^2 \, \fq \, y + 2 \, a_0 \, a \, b^2 \, \fq \, y - 2 \, a_0 \, a \, b^2 \, \fq^2 \, y + 
 a_0^2 \, b^2 \, y^2 \notag \\
 & - & 2 \, a_0^2 \, b^2 \, \fq \, y^2 + a_0^2 \, b^2 \, \fq^2 \, y^2 + 2 \, a_0 \, b \, z  - 2 \, a_0^2 \, b \, z - 2 \, a_0 \, b^2 \, z 
 + 2 \, a \, b \, \fq \, z - 4 \, a_0 \, a \, b \, \fq \, z + 2 \, a_0^2 \, a \, b \, \fq \, z \label{defQ1} \\
 & - & 2 \, a \, b^2 \, \fq \, z + 4 \, a_0 \, a \, b^2 \, \fq \, z - 2 \, a^2 \, b \, \fq^2 \, z + 2 \, a_0 \, a^2 \, b \, \fq^2 \, z 
 - 2 \, a_0 \, b \, y \, z + 2 \, a_0^2 \, b \, y \, z + 2 \, a_0 \, b^2 \, y \, z + 2 \, a_0 \, b \, \fq \, y \, z \notag \\
 & - & 2 \, a_0^2 \, b \, \fq \, y \, z - 2 \, a_0 \, a \, b \, \fq \, y \, z + 2 \, a_0^2 \, a \, b \, \fq \, y \, z - 2 \, a_0 \, b^2 \, \fq \, y \, z
 + 2 \, a_0 \, a \, b \, \fq^2 \, y \, z - 2 \, a_0^2 \, a \, b \, \fq^2 \, y \, z + z^2 - 2 \, a_0 \, z^2 \notag \\
&+& a_0^2 \, z^2 -2 \, b \, z^2 + 2 \, a_0 \, b \, z^2 + b^2 \, z^2 - 2 \, a \, \fq \, z^2 + 4 \, a_0 \, a \, \fq \, z^2 - 
 2 \, a_0^2 \, a \, \fq \, z^2 + 2 \, a \, b \, \fq \, z^2 \notag \\
 &-& 2 \, a_0 \, a \, b \, \fq \, z^2 + a^2 \, \fq^2 \, z^2 - 2 \, a_0 \, a^2 \, \fq^2 \, z^2 + a_0^2 \, a^2 \, \fq^2 \, z^2 \, . \notag
\end{eqnarray}
Thanks to this property the problem will stay quadratic also at the second level: in each step it is the discriminant of the quadratic polynomial that will progress to the next. Our analysis here is in fact a slightly more complicated version of the earlier investigation \cite{usInter} that also showed this feature. 

With \eqref{basS} as the left and right basis the matrix of $\mathbf{s}$ intersection numbers equals
\beq
C_\fs \eqsp \frac{1}{4 \, \gamma} \, D_\fs^{-1} . \left( \begin{array}{ccc} 3 & 1 & 0 \\ 1 & 3 & 0 \\ 0 & 0 & 8 \end{array} \right) . \, D_\fs^{-1} \, , \qquad D_\fs \eqsp \mathrm{diag} \left\{ \fq, \, \fq, \, y \, \sqrt{Q_1} \right\}  \, . \label{Cs}
\eeq
Remarkably, $C_\fs$ is homogeneous of order $1/\gamma$ and the coordinate dependence\footnote{The parameter $\fq$ is an outer variable from the point of view of the $s$ integration.} factors out! What is more, the purely numerical part of $C_\fs$ has determinant 1. For bases of masters the matrix of intersection numbers should indeed always be invertible. This implies
\beq
\sum_{j \eqsp1}^3 \la \phi_L | h_j \ra (C_\fs^{-1})_{ji} \eqsp \phi'_L(\fq)_i \, \text{d}\fq \, . \label{decS}
\eeq
Henceforth we will refer to the $\phi_L'$ as \emph{$\fq$-projected left forms}. In our example they are three-component objects. We will need $\fq$-projected right masters, too, though without an explicit left multiplication by $C_\fs^{-1}$.  Let us introduce the notation
\beq
\varphi'_R(\fq)_i \, \text{d}\fq \eqsp \la e_i | \phi_R \ra \, , \qquad \phi_R' \eqsp C_\fs^{-1} \varphi'_R \, .
\eeq
The connection for the second reduction step is defined by the covariant derivative of the level 1 basis \eqref{basS} \cite{inter2}: let
\beq
\nabla_\fq \eqsp \partial_\fq + \hat \omega_\fq
\eeq
and define
\beq
(\Omega_\fq)_{ij} \eqsp \sum_{k \eqsp 1}^3 \la \nabla_\fq e_i | h_k \ra (C_\fs^{-1})_{kj} \, . \label{makeOmT}
\eeq
where the $\fs$ intersection number in the angle brackets is constructed just as described for the left forms themselves, i.e. building a potential and then taking residues in its product with the right level 1 masters. This connection ought to have poles of maximally first order in $\fq$ which are the singularities to expand around at level 2. These are the zeroes of
\beq
\left\{\fq, \, 1 - \fq, \, P_3, \, P_4, \, P_5, \, P_6, \, Q_1, \, \frac{1}{\fq} \right\}
\eeq
where
\beq
P_5 \eqsp 1 - b - a \, \fq - a_0 \, y + a_0 \, \fq \, y \, , \qquad P_6 \eqsp 1 - a \, \fq - a_0 \, y + a_0 \, \fq \, y - z + a \, \fq \,  z \, .
\eeq
In fact, $P_1|_{\fs \rar 1} \eqsp P_3 \, P_4$, which is why these polynomials also arose from partial fractions of $P_1$ with powers of $1-\fs$. Likewise, $P_1|_{\fs \rar 0} \eqsp P_5 \, P_6$. The pole from $P_1|_{\fs \rar \infty}$ is $c_2 \eqsp (1-a_0)(1-a_0+b) (1-\fq)^2 \, y^2$ and so does not contribute anything new.  By construction, the set of poles $\{\fq_1 \ldots \fq_9\}$ found in the $\fq$-projected left or right quantities is covered.

We will replace $\fq \rar \fq_m + \fw$ and expand in the new parameter $\fw$. For the $\fq$ intersection numbers we need a potential $\psi$, too, of which a range of powers $min \ldots max$ in $w$ is needed. It must be a three-component object so that we impose a matrix version of equation \eqref{defPsiV}:
\beq
\partial_\fw \, \psi  + \psi \, . \, \Omega_\fq|_{\fq \eqsp \fq_m + \fw} \eqsp \hat \phi'_L|_{\fq \eqsp \fq_m + \fw} \label{defPsiW}
\eeq
where $\phi'_L$ on the r.h.s. is also the collection of all three components. Their lowest order $min_m$ determines the start of the expansion of the potential, but to be on the safe side we should take the absolute minimum of the lowest orders  estimates of the three components of $\phi'_L$, which we call $min_m-1$ again since $\Omega_\fq$ as well as the partial $\fq$ derivative lower the order in $w$ by 1. On the other hand the maximum order $max_m$ will be defined by the $\fq$-projected right master in question, taking the absolute maximum of the three order estimates $-min_m-1$ for the  individual components.

The level 2 intersection number is
\beq
C_\fq \eqsp \la \phi_L | \phi_R \ra_\fq \eqsp \sum_{m \eqsp_1}^l res|_{\fq \eqsp \fq_m} \, \psi \, . \, \varphi_R'|_{\fq \eqsp \fq_m +\fw} \label{defCq}
\eeq
with $\psi$ defined by \eqref{defPsiW}. Here $\phi_L$ and $\phi_R$ are members of larger bases completing the level 1 choice. 

A comment is due: at the various orders of the series in $\fw$ the operator $\partial + \Omega$ in \eqref{defPsiW} may yield matrices that are not invertible. Remarkably, the equations \eqref{defPsiW} always have a solution, and any homogeneous part will be in the kernel of the dot product on any $\fq$-projected right master. The intersection numbers \eqref{defCq} therefore exist and are well-defined.

Yet, in the \emph{root regions} where we expand around the two zeroes $\fq^\mp$ of $Q_1$ in \eqref{defQ1} this presents a coding problem because {\tt Mathematica} may fail to recognise the equivalence of various writings of the same expression and therefore be unable to solve the linear equations for the vector $\psi$. Choosing bases of masters with maximally simple poles in $P_1$ substantially helps: the differentiation wrt. the kinematic variables $\{a_0, \, a, \, b, \, y, \, z\}$ may increase this to a second order pole. We have also been careful in the last section to avoid poles of more than second order in the integrals we want to reduce. As a consequence, the $\fq$-projected left forms will have poles of maximally second order in $Q_1$ prior to right multiplication by $C_\fs^{-1}$. Since the latter has a factor $Q_1$ in the bottom right corner, undifferentiated left masters will actually be free of $Q_1$ poles, while differentiated ones and the three integrals of interest displaying a double pole in $P_1$ will have a first order pole in $Q_1$ after taking $\fs$ intersections. Similarly, the $\fq$-projected right masters have at maximum a first order pole in $Q_1$. It follows that we only need the constant order of $\psi$ at $\fq^\pm$. In most cases the contribution of the root regions simply vanishes, eg. they do not touch upon $C_\fq$ itself.

Next, in these regions the connection has the form
\beq
\Omega_\fq|_{\fq^\mp} \eqsp \frac{1}{w} \left( \begin{array}{ccc} 0 & 0 & 0 \\ 0  & 0 & 0 \\ * & * & \gamma - \frac{1}{2} \end{array} \right) + \ldots
\eeq
Multiplying this from the right onto $\psi_i^{(0)}$ selects only the component $\psi_3^{(0)}$ so that equating on any $\fq$-projected right form constitutes an overdetermined system. It is clearly most convenient to solve the third equation. We can verify the consistency of the other two conditions reducing to simple roots of the discriminant\footnote{$\Delta_\fq \eqsp 16 \, a_0 \, a \, b \, z  \, (1-a_0) \,  (1- a_0 - b) \, (b-z) \, D_{10}$, cf. \eqref{D1011}}  $\Delta_\fq$ of $Q_1$ and eliminating the latter from the denominator using the third Binomi formula. These elementary ideas worked in a range of problems with multi-quadratic denominators, also for more than two integration steps.
 
According to \eqref{countMs} the basis length for the full problem is 13. We choose
\beq
\begin{split}
&\left\{ \frac{1}{P_1}, \frac{1-\fq}{\fq \, P_1}, \frac{1}{(1-\fs) \,  P_1}, \frac{1}{\fs \, P_1}, \frac{1}{(1-\fq) \, (1-\fs)}, \frac{1}{(1-\fq) \,  \fs},\right.\\&
 \left.\frac{1}{(1-\fs) \,  P_3}, \frac{1}{\fs \, P_3}, \frac{1}{(1-\fs) \,  P_4}, \frac{1}{\fs \, P_4}, \frac{1}{\fq \, \fs}, \frac{1}{\fq \, (1-\fs)}, \frac{1-\fq}{P_1} \right\}
\end{split}
 \label{willi}
\eeq
for both, the left and right basis. The last three elements are the level 1 masters. This choice is convenient but not mandatory.

Like in our analysis of the root regions, it is quite generally advisable to allow only simple poles in the forms in the basis; according to the order estimates we will then only need the lowest order of the potentials to determine $C_\fq$. Since we want to freeze both integrations, the obvious choice is $1/(\fq \, \fs)$ and similar; in \eqref{willi} all basis elements without $P_1$ are of this type. The irreducible polynomial $P_1$ itself can serve as a denominator on its own because it has simple poles in $\fs$ in the first step and will be replaced by its discriminant in the second step (plus simpler terms), whereas $Q_1$ has once again simple poles in $\fq$.  By way of example, the denominator $(1-\fs) P_1$ reacts to leading order like $(1-\fs) \, P_3 \, P_4$: recall that freezing $\fs$ at $1$ makes $P_1$ factor into $P_3 \, P_4$.
 
We find the second level intersection numbers
\beq 
C_\fq \eqsp \frac{1}{24 \, \gamma^2} \, D_\fq^{-1} . \left(
\begin{array}{rrrrrrrrrrrrr}
 48 & 0 & 0 & 0 & 0 & 0 & 0 & 0 & 0 & 0 & 0 & 0 & 0 \\
 0 & -48 & 0 & 0 & 0 & 0 & 0 & 0 & 0 & 0 & 0 & 0 & 0 \\
 0 & 0 & -32 & 0 & 0 & 0 & 8 & 0 & -8 & 0 & 0 & 0 & 0 \\
 0 & 0 & 0 & -48 & 0 & 0 & 0 & 0 & 0 & 0 & 0 & 0 & 0 \\
 0 & 0 & 0 & 0 & -3 & -17 & 3 & 1 & 3 & 1 & -1 & -3 & 0 \\
 0 & 0 & 0 & 0 & -17 & -3 & 1 & 3 & 1 & 3 & -3 & -1 & 0 \\
 0 & 0 & 8 & 0 & 3 & 1 & -7 & -5 & 3 & 1 & -1 & -3 & 0 \\
 0 & 0 & 0 & 0 & 1 & 3 & -5 & -15 & 1 & 3 & -3 & -1 & 0 \\
 0 & 0 & -8 & 0 & 3 & 1 & 3 & 1 & -7 & -5 & -1 & -3 & 0 \\
 0 & 0 & 0 & 0 & 1 & 3 & 1 & 3 & -5 & -15 & -3 & -1 & 0 \\
 0 & 0 & 0 & 0 & -1 & -3 & -1 & -3 & -1 & -3 & -15 & -5 & 0 \\
 0 & 0 & 0 & 0 & -3 & -1 & -3 & -1 & -3 & -1 & -5 & -15 & 0 \\
 0 & 0 & 0 & 0 & 0 & 0 & 0 & 0 & 0 & 0 & 0 & 0 & 8 \\
\end{array}
\right) . \, D_\fq^{-1} \label{Cq}
\eeq
with
\beq
D_\fq \eqsp  \mathrm{diag}\left( \{y \, \sqrt{D_9} , \, y \, D_6, \, D_4, \, D_7, \, 1, \, 1, \, a - y, \, a - y, \, D_1, \, D_1, \, 1, \, 1, \, y \, D_8\} \right) \, . \label{densSQ}
\eeq
In the last formula
\begin{eqnarray}
D_6 & = & a_0 \, b - a_0 \, b \,  y + z - a_0 \, z - b \,  z \, , \notag \\
D_7 & = & a \,  b - a_0 \, b \,  y - a \,  b \,  z + a_0 \, y \,  z - a_0 \, a \,  y \,  z \, , \notag \\
D_8 & = & a \, b - a_0 \, b \, y - a \, z + a_0 \, a \, z \, , \\
D_9 & = & a_0^2 \, b^2 - 2 \, a_0 \, a \, b^2 + a^2 \, b^2 + 2 \, a_0 \, b \, z - 2 \, a_0^2 \, b \, z + 
 2 \, a \, b \, z - 4 \, a_0 \, a \, b \, z + 2 \, a_0^2 \, a \, b \, z - 2 \, a^2 \, b \, z + 
 2 \, a_0 \, a^2 \, b \, z \, , \notag \\
 &-& 2 \, a_0 \, b^2 \, z -2 \, a \, b^2 \, z + 4 \, a_0 \, a \, b^2 \, z + z^2 - 2 \, a_0 \, z^2 + a_0^2 \, z^2 - 2 \, a \, z^2 + 
 4 \, a_0 \, a \, z^2 - 2 \, a_0^2 \, a \, z^2 + a^2 \, z^2 \notag \\
 &-& 2 \, a_0 \, a^2 \, z^2 + a_0^2 \, a^2 \, z^2 - 2 \, b \, z^2 + 2 \, a_0 \, b \, z^2 + 2 \, a \, b \, z^2 - 2 \, a_0 \, a \, b \, z^2 + b^2 \, z^2 \, . \notag
 \end{eqnarray}
The entire problem is in fact a five-variable version of the $\hat I_1$ analysis of \cite{usInter}: up to a slight re-ordering of the basis  it reduces to the one analysed there under the replacement $a_0 \rar a$. The two non-boundary denominators without $\fs$ are even identical with $P_3, \, P_4$ here. The form \eqref{Cs}, \eqref{Cq} of the intersection numbers $C_\fs, \, C_\fq$ is pivotal for success with the method, see also \cite{usInter}: if the matrices had a complicated coordinate dependence, their determinant as a  denominator of the inverses would hinder any progress with the decompositions \eqref{decS} and at the second level \eqref{decQ} below. With the \emph{dLog} bases \eqref{basS}, \eqref{willi} the inverses will compensate existing denominator factors instead of introducing new and very involved ones.
   
Finally, any two-form can be decomposed in terms of the second level left basis using
\beq
\phi \eqsp \la \phi \, | \, \underline h \, \ra \, C_\fq^{-1} \, , \label{decQ}
\eeq
where $\underline h$ is the full level 2 right basis. In particular, this is so for derivatives of the left masters wrt. outer parameters. Let
\beq
M_x \eqsp \la (\partial_x + \hat \omega_x) \, \underline{e} \, | \, \underline{h} \, \ra \, C_\fq^{-1} \, , \qquad \hat \omega_x \eqsp \partial_x \, \log(u) \, .
\eeq 
This allows us to write one Pfaffian differential equation
\beq
\partial_x \, \underline{e} \eqsp M_x \, . \, \underline{e} \label{doPfaff}
\eeq
for every outer parameter.

The Pfaffian matrices are too involved to be displayed here. Nonetheless, they are linear in $\gamma$. The constant order can be removed multiplying the basis \eqref{willi} by $D_\fq$ from \eqref{densSQ} upon which the Pfaffian matrices will be proportional to $\gamma$. We obtain a set of five coupled \emph{canonical} equations in the sense of \cite{hennAlg}. In the canonical form the $O(\gamma)$ matrices commute so that the equations can be solved variable by variable.

Furthermore, they can be solved iteratively: singularities of the basis \eqref{willi} arise from divergences at the boundaries $\fs \eqsp 0, \, 1$ and $\fq \eqsp 0, \, 1$ of the integration domain, cf. \cite{usInter}. To leading order the integrals of the entries $\# 5, \, 6, \, 11, \, 12$ of the basis \eqref{willi} are double Euler $\beta$ functions, and so the highest poles are
\beq
v^{(-2)} \eqsp \frac{1}{\gamma^2} \{ 0, \, 0, \, 0, \, 0, \, 1, \, 1,\, 0, \, 0, \, 0, \, 1, \, 1, \, 0 \} \label{lowestGamma}
\eeq
To first order, the master integrals are given as
\beq
v^{(-1)}_{a_0} \eqsp \int da_0 \, M_{a_0} \, . \,  v^{(-2)} 
\eeq
which is submitted into the next equation, in which the variable $a_0$ drops out and so on. Hence at $O(\gamma^{-1})$ the masters will receive logarithmic contributions. This concerns all elements barring for the first and the last, which are finite. We are obviously building up iterated integrals; more specifically Goncharov polylogarithms \cite{symbolGonchiLogs}. For simplicity, we have been using {\tt Mathematica}'s inbuilt integration throughout this work because there are no non-classical polylogarithms up to weight two.

Where the logarithmic order is leading, the result can be determined also by \emph{1d unitarity}: consider for instance the fourth master that we had already discussed above. The $\fs$ integration is divergent at $\fs \eqsp 1$ bringing out a factor $1/\gamma$. Now, freezing the rest of the integrand at the singularity:
\begin{eqnarray}
\iint_0^1 \frac{\text{d}\fs \, \text{d}\fq}{(1-\fs) \, P_1} & = & \frac{1}{\gamma} \int_0^1 \frac{\text{d}\fq}{P_3 \, P_4}  + \ldots  \label{exampleAs} \\
& = & \frac{\log[1 - a] + \log[1 - b] - \log[1 - a - b] - \log[1 - y] - \log[1 - z] + \log[1 - y - z]}{\gamma \, D_4} + \ldots \notag
\end{eqnarray}
Scaling up by the fourth entry of $D_\fq$ (cf. \eqref{densSQ}) we obtain a \emph{pure function}, so a combination of logarithmic functions without rational coefficients.

At $O(\gamma^{-1})$ the differential equations contain only multilinear denominators, and these do only occur to first order. Integration is thus trivial and the solution of the differential equations does reproduce estimates like \eqref{exampleAs}. In addition, it also yields logarithmic corrections to the master integrals $\# 5, \, 6, \, 11, \, 12$ which we could not derive by simple means. Similarly, from the asymptotics we cannot easily deduce any knowledge about the leading term of the finite integrals $\# 1, \, 13$. In general, at $O(\gamma^0)$ we expect to find dilogarithms, but we will have to use the Pfaffian equations to derive these. 

Yet, the root denominator of the first master seems an obstacle to integration. Luckily, closer inspection of the decomposition of the six integrals from the last section shows that the first master never contributes! In fact, we only need to integrate masters $\# 3, \, 7, \, 9$ up to dilogarithm level. Rows $\# 3, \, 7, \, 9$ of the five Pfaffian matrices have a range of multilinear denominator polynomials, but also the combinations
\begin{eqnarray}
D_{10} & = & a \, b - a \, b \, y - a_0 \, b \, y + a_0 \, b \, y^2 - a \, b \, z - y \, z + a \, y \, z + a_0 \, y \, z - a_0 \, a \, y \, z + b \, y \, z \, , \label{D1011} \\ 
D_{11} & = & a \, b - a_0 \, a \, b - a \, b^2 - a \, b \, y + a_0 \, a \, b \, y + a \, b^2 \, y - a \, b \, z + 
  2 \, a_0 \, a \, b \, z + a \, b^2 \, z - y \, z + a \, y \, z \notag \\
 &+& a_0 \, y \, z - a_0 \, a \, y \, z + 2 \, b \, y \, z -  a \, b \, y \, z - 2 \, a_0 \, b \, y \, z - b^2 \, y \, z + a_0 \, b^2 \, y \, z - a_0 \, a \, b \, z^2 \, , \notag
\end{eqnarray}
which are quadratic in $y$ and $b, \, z$, respectively. As mentioned before, integration will remain a linear problem as long as we start with the $a_0$ or $a$ equation. Either one can be used to soak up all quadratic dependence in the very first step, after which the other partial differential equations will not contain $D_{10}, \, D_{11}$ anymore and so become elementary to solve, too. Uniform transcendentality is manifest because all denominators occur only to first order.

The six integrals in question do not have singularities in $\gamma$. Combining the reduction of the two B sector three-parameter integrals, the two A sector three-parameter integrals, and the two A sector four-parameter integrals, respectively, the results for the finite parts are rational and logarithmic in the first case, and they have a rational, logarithmic and dilogarithmic part in the second and third case. Thus unlike the master integrals themselves they are not of uniform transcendentality \cite{lipKot}. What is more, the dilogarithm parts are given by some rational factor multiplying a pure function. Curiously, these two pure functions are equal although the latter property is obviously destroyed upon scaling back to the original variables. We include the results of the integral reduction in the ancillary material.

\section{Remaining residues}

For the calculation of the remaining residues, we will not work through the re-summation in detail as it follows the exact same scheme as the $\mathcal{Z}_{33}$-polygamma residue from the previous section. Instead we provide a general guideline through each residue-type. For the evaluation of the hypergeometric function we use the following rules, which are special cases of \eqref{reducePFP}:

\subsubsection*{Integral representation of $_4F_3$}
\begin{equation}
\label{eq:rule4F3I}
\begin{split}
&_4F_3[\{a_1, a_2, a_3, a_4\}, \{b_1, b_2, b_3\};\, z]=\\
&\frac{ \Gamma [b_3]}{\Gamma [a_4] \Gamma [b_3-a_4] }\int_0^1 \mathbf{x}^{a_4-1}\, (1-\mathbf{x})^{b_3-a_4-1}\, _3F_2[\{a_1,a_2, a_3\},  \{b_1, b_2\};\, z\,\mathbf{x}] \text{d}\mathbf{x}
\end{split}
\end{equation}
\subsubsection*{Integral representation of $_3F_2$}
\begin{subequations}
\begin{equation}
\label{eq:rule3F2a1b1}
_3F_2[\{a_1, a_2, a_3\}, \{b_1, b_2\};\, z]=\frac{ \Gamma [b_1]}{\Gamma [a_1] \Gamma [b_1-a_1] }\int_0^1 \mathbf{x}^{a_1-1}\, (1-\mathbf{x})^{b_1-a_1-1}\, _2F_1[\{a_2, a_3\}, b_2;\, z\,\mathbf{x}] \text{d}\mathbf{x}
\end{equation}
\begin{equation}
\label{eq:rule3F2a3b2}
_3F_2[\{a_1, a_2, a_3\}, \{b_1, b_2\};\, z]=\frac{ \Gamma [b_2]}{\Gamma [a_3] \Gamma [b_2-a_3] }\int_0^1 \mathbf{x}^{a_3-1}\, (1-\mathbf{x})^{b_2-a_3-1}\, _2F_1[\{a_1, a_2\}, b_1;\, z\,\mathbf{x}] \text{d}\mathbf{x}
\end{equation}
\end{subequations}
\subsubsection*{Integral representation of $_2F_1$}
\begin{subequations}
\begin{equation}
_2F_1[\{a_1, a_2\}, b_1;\, z]=\frac{ \Gamma[b_1]}{\Gamma[a_2] \Gamma[b_1-a_2]} \int_0^1 (1-\mathbf{x})^{-1-a_2+b_1} \mathbf{x}^{a_2-1} (1-\mathbf{x} z)^{-a_1} \text{d}\mathbf{x}
\label{eq:rule2F1I}
\end{equation}
\begin{equation}
_2F_1[\{a_1, a_2\}, b_1;\, z]=\frac{ \Gamma[b_1]}{\Gamma[a_1] \Gamma[b_1-a_1]} \int_0^1 (1-\mathbf{x})^{-1-a_1+b_1} \mathbf{x}^{a_1-1} (1-\mathbf{x} z)^{-a_2} \text{d}\mathbf{x}
\label{eq:rule2F1II}
\end{equation}
\end{subequations}

\subsection{Re-summation of $S(\mathcal{M})^u_{\epsilon}$}

\label{ssec:UX}
Starting out with the residue of type $u\to i(K/2 +\epsilon)$ where $\epsilon$ is a regulator, it is worth noting that the contribution $\mathcal{S}(\mathcal{Y}_{11})^u_{\epsilon}$, $\mathcal{S}(\mathcal{Y}_{33})^u_{\epsilon}$,  and $\mathcal{S}(\mathcal{Y}_{44})^u_{\epsilon}$  take the general form
\begin{equation}
S^u_{\epsilon}=\sum\frac{{K+k+m}}{{K+k+m+\epsilon}} \frac{f(\epsilon)}{\epsilon}.
\label{eq:splitSU}
\end{equation}
Here $f(\epsilon)$ is completely regular in $\epsilon$ and contains all remaining pieces of the full expression. The pre-factor of $f$ labels the derivative acting on the measure and we can simplify $S^u_{\epsilon}$ by the separation
\begin{equation}
\left(\frac{{K+k+m}}{{K+k+m+\epsilon}}\right)\frac{f(\epsilon)}{\epsilon}=\left(1-\frac{\epsilon}{{K+k+m}}+O\left(\epsilon^2\right) \right)\frac{f(\epsilon)}{\epsilon}=\frac{f(\epsilon)}{\epsilon}-\frac{1}{{K+k+m}}f(\epsilon=0).
\label{eq:separatUX}
\end{equation}

The last step uses the fact that $f(\epsilon)$ in itself does not contain any divergences with respect to $\epsilon$. As a result we can split \eqref{eq:splitSU} as
\begin{equation}
S^u_{\epsilon}=S^u_{\epsilon,\,\text{main}}+S^u_{\epsilon, \text{mes}} \, .
\end{equation}
The $S^u_{\epsilon, \text{mes}}$-term follows the same re-summation steps as the main term and is left out of the detailed discussion\footnote{Additionally, its explicit dilogarithm result was also already stated for all cases in \cite{usFivePoints}}.  That being said, the evaluation of the $S^u_{\epsilon,\,\text{main}}$ is by far the simplest compared to the other residues. Consequently, we are able to tackle the series in various different orders. However, it is best to pick $l$ first, followed by $L$, $k$, $K$ and finally $m$. This is true for all matrix elements and the only difference is the number of Euler integrals created in each summation step. For instance, starting on the $l$-series will either result in a $_3F_2$ or a $_2F_1$. 
The second sum over $L$ is always of geometric type while the $k$-series can be geometric or result in $_{p+1}F_p$ with $p \, \in \, \{1,2,3\}$. Finally, the $K$ and $m$ series are always geometric. Similarly to the complicated problem presented in section \ref{hardest}, we find a common denominator for all contributions. It takes the general form
   \begin{equation}
   d_{S^u_{\epsilon}}=1- a\,\mathbf{q}- b\,\mathbf{s}-y+ y\,\mathbf{q}+ b\, y\,\mathbf{s}- b\, y\,\mathbf{q}\,\mathbf{s}- z\,\mathbf{s}+ a\, z\,\mathbf{q}\,\mathbf{s}+ b \,z\,\mathbf{s}^2
   \label{eq:DefDenUX}
   \end{equation}
with the following set of transformations for re-scaling cross ratios:
\begin{subequations}
\begin{align}
s_0&=\{ a\rightarrow a, b\rightarrow b, y\rightarrow y, z\rightarrow z\},\\
s_{r, s \to 1}&=\{ a\rightarrow a\,\mathbf{r}, b\rightarrow b, y\rightarrow y\,\mathbf{r}, z\rightarrow z; (s\rightarrow1)\},\\
s_{r^2, s \to 1}&=\{ a\rightarrow a\,\mathbf{r}, b\rightarrow b\,\mathbf{r}, y\rightarrow y\,\mathbf{r}, z\rightarrow z\,\mathbf{r}; (s\rightarrow1)\},\\
s_{r}&=\{ a\rightarrow a\,\mathbf{r}, b\rightarrow b, y\rightarrow y\,\mathbf{r}, z\rightarrow z\},\\
s_{r^2,t}&=\{ a\rightarrow a\,\mathbf{r}, b\rightarrow b\,\mathbf{r}\,\mathbf{t}, y\rightarrow y\,\mathbf{r}, z\rightarrow z\,\mathbf{r}\,\mathbf{t}\},\\
s_{r,t}&=\{ a\rightarrow a\,\mathbf{r}, b\rightarrow b\,\mathbf{t}, y\rightarrow y\,\mathbf{r}, z\rightarrow z\,\mathbf{t}\}.
\end{align}   
\end{subequations}
The parameters $\mathbf{q}$ and $\mathbf{r}$ are always obtained in the summations over $l$. The different kinds of $\mathbf{r}$ re-scalings are caused by the two different denominators ${\epsilon+ l+ L+ 2 \, m} $ or ${\epsilon +k +K +l+ L +3 \, m}$. This is the direct analogue of the B and A type terms in Section \ref{hardest}. For the first case we always use \eqref{eq:rule3F2a1b1} and \eqref{eq:rule2F1II} to transform $_3F_2$ into a double integral over $\mathbf{q}$ and $\mathbf{r}$ while in the second case \eqref{eq:rule3F2a1b1} and \eqref{eq:rule2F1I} are employed. Should the $k$-series result in a hypergeometric function, a $_3F_2$ is translated into an integration over $\mathbf{s}$ and  $\mathbf{t}$ using  \eqref{eq:rule3F2a3b2} and \eqref{eq:rule2F1I}. Any $_2F_1$ is expressed as an integral over $\mathbf{s}$ with  \eqref{eq:rule2F1I}. In rare cases a $_4F_3$-function is obtained. Yet, this will always take the form $_4F_3(\{2,...\};\{1,...\};...)$. Via \eqref{eq:rule4F3I} and \eqref{eq:rule3F2a3b2} it can be decomposed into a double integral in $\mathbf{s}$ and $\mathbf{t}$ over a remaining $_2F_1$. Due to the special indices $2,1$ the latter reduces to a geometric sum so that a third integral does not arise.

Tab.: \ref{tab:ScaleUReg} illustrates which scalings are are found in the various diagonal contributions:
\begin{table}[h]
\begin{center}
\begin{tabular}{ c|c|c|c|c|c|c|c } 
 $ $ & 	$s_{0}$ &$s_{\mathbf{r}, \mathbf{s} \to 1}$& $s_{\mathbf{r}^2, \mathbf{s} \to 1}$& $s_{\mathbf{r}}$& $s_{\mathbf{r}^2}$  & $s_{\mathbf{r}^2,\mathbf{t}}$ & $s_{\mathbf{r},\mathbf{t}}$ \\ 
  \hline
   \hline
 $\mathcal{Y}_{11}$ & $\star$ &$ $& $ $ & $ $ & $ $& $ $&  $ $  \\ 
 $\mathcal{Y}_{33}$ & $ $ &$ \star$& $\star$&  $\star$& $\star $ &  $ $& $ $  \\ 
 $\mathcal{Y}_{44}$ & $ $ &$ $& $\star$& $\star$ & $ $& $ $& $ $   \\ 
 $\mathcal{Z}_{33}$ & $ $ &$ $& $$ & $\star$ & $ \star$&$\star $ & $\star$   \\ 
 $\mathcal{Z}_{44}$ & $ $ &$ $& $ $& $$& $ \star $ & $\star$& $\star$  \\
 \hline
\end{tabular}
\end{center}
\caption{Rescaling rules for the cross-ratios for the $S^u_{\epsilon}$-residues}
\label{tab:ScaleUReg}
\end{table}

 \subsubsection*{Integration} 
 
 After the re-summation is complete, it is safe to expand the resulting Euler integrals in $\epsilon$ up to $O(\epsilon^0)$. All denominators but $d_{S_\epsilon^u}$ spelled out in \eqref{eq:DefDenUX} are multilinear in the Euler parameters. Given that $d_{S_\epsilon^u}$ is quadratic in $\fs$ one will rather integrate in $\mathbf{q}$ first, upon which the $\fs$ dependence linearises, too. The numerators of the integrals drastically differ in terms of complexity but they present no obstacle to integration.

\subsection{Re-summation of $S(\mathcal{M})^u_{\psi}$}

The next case to consider will be the residue with $u \to i(K/2 +j)$ where $j$ acts as a summation index. Since $j$ is not a regulator, it does not help to isolate the measure contribution as we did in \eqref{eq:splitSU}. We can take the summation in different orders: $l,L,j,K,k,m$ or  $l,k,L,j,K,m$ and $l,L,k,K,m,j$ are all valid choices. 
In general, the last option seems most convenient:  as in \ref{ssec:UX}, summing over $l$ results in either a $_3F_2$ or a $_2F_1$. While $L$ is a simple geometric series, the $k$ summation is either of geometric or $_2F_1$ type. This is followed by two geometric limits from the sum over $K$ and $m$. Finally, the summation over $j$ can be of geometric or logarithmic type. The defining common denominator is the same as for the previous case:
      \begin{equation}
   d_{S^u_{\psi}}=1- a\,\mathbf{q}- b\,\mathbf{s}-y+ y\,\mathbf{q}+ b \,y\,\mathbf{s}- b\, y\,\mathbf{q}\,\mathbf{s}- z\,\mathbf{s}+ a\, z\,\mathbf{q}\,\mathbf{s}+ b \,z\,\mathbf{s}^2
   \label{eq:DefDenUJ}.
   \end{equation}
In total, four rescaling rules ar found for \eqref{eq:DefDenUJ}:
\begin{subequations}
\begin{align}
s_{\mathbf{r}}&=\{ a\rightarrow a\,\mathbf{r}, b\rightarrow b, y\rightarrow y\,\mathbf{r}, z\rightarrow z\}, \\
s_{\mathbf{r},\mathbf{s}\to1}&=\{ a\rightarrow a\,\mathbf{r}, b\rightarrow b, y\rightarrow y\,\mathbf{r}, z\rightarrow z; \mathbf{s}\rightarrow1\} ,\\
s_{\mathbf{r}^2}&=\{ a\rightarrow a\,\mathbf{r}, b\rightarrow b\,\mathbf{r}, y\rightarrow y\,\mathbf{r}, z\rightarrow z\,\mathbf{r}\},\\
s_{\mathbf{r}^2,\mathbf{s}\to1}&=\{ a\rightarrow a\,\mathbf{r}, b\rightarrow b\,\mathbf{r}, y\rightarrow y\,\mathbf{r}, z\rightarrow z\,\mathbf{r};\mathbf{s}\to1\}.
\end{align}
\end{subequations}
For all  the concerned matrix elements, the re-summation of this series of residues runs up at most three Euler parameters. The two integrations $\mathbf{q}$ and $\mathbf{r}$ are obtained in the exact same way as the previous case sketched in Section \ref{ssec:UX}. If an integration over $\mathbf{s}$ is present, it is due to using \eqref{eq:rule2F1I} in the summation over $k$. 

Table \ref{tab:rescalingsUJ} lists which re-scalings were used in calculating the $S({\cal M}_\psi^u)$ contribution of what matrix element:
\begin{table}[h]
\begin{center}
\begin{tabular}{ c|c|c|c|c } 

 $ $ & 	$s_{\mathbf{r}}$ & $s_{\mathbf{r}^2}$& $s_{\mathbf{r}, \mathbf{s} \to 1}$  & $s_{\mathbf{r}^2, \mathbf{s} \to 1}$  \\ 
  \hline
   \hline
 $\mathcal{Y}_{22}$ & $\star$ & $ $ &  $\star$ & $ $ \\ 
 $\mathcal{Y}_{33}$ & $\star $ & $\star$&  $ $ &  $ $  \\ 
 $\mathcal{Y}_{44}$ & $ $ & $\star$& $ $ & $\star$   \\ 
 $\mathcal{Z}_{22}$ & $\star$  & $\star$ & $ $ & $\star$   \\ 
 $\mathcal{Z}_{44}$ & $\star$  & $\star $& $ $ & $\star$ \\
 \hline
\end{tabular}
\end{center}
\caption{Rescaling rules for the cross-ratios for the $S^u_{\psi}$-residues}
\label{tab:rescalingsUJ}
\end{table}

Furthermore, if a logarithm is obtained in evaluating the $j$ sum, it always takes the rescaling-independent form
\begin{equation}
\text{Log}\left[\frac{1-a\, \mathbf{q} \mathbf{r}-y\,\mathbf{r} +y\,\mathbf{q} \mathbf{r} }{1-a\, \mathbf{q} \mathbf{r}-y\,\mathbf{r} +y\,\mathbf{q} \mathbf{r} -b\,z\, \mathbf{r} }\right].
\end{equation}

 \subsubsection*{Integration}  
 
As the integration does not contain any regulator we are free to tackle the integration in $\mathbf{q}$, $\mathbf{r}$ and $\mathbf{s}$ right away. However, as explained in section \ref{ssec:UX}, the form of the common denominator suggests to start the evaluation with $\mathbf{q}$ or --- if present --- $\mathbf{s}$. 
Subsequent integrations are rather trivial for the three $\mathcal{Y}$ elements, but become more complicated for the two $\mathcal{Z}$-elements because logarithms with square root arguments may appear in the last integration step. However, such terms do not cause any trouble during the integration itself and they do eventually cancel non-trivially. Furthermore, trilogarithms, which may arise during the integration process, will also cancel non-trivially in the very end.

\subsection{Re-summation of $S(\mathcal{M})^v_{\epsilon}$}

We are now going to study the residue $v\to-i(L/2+\epsilon)$ with $\epsilon$ acting as a regulator. For $\mathcal{Y}_{22}$, $\mathcal{Y}_{33}$ and $\mathcal{Y}_{44}$ we observe the common structure
\begin{equation}
S^{v}_{\epsilon}=\sum\frac{{l +L +2 \, m}}{{\epsilon +l +L +2 \, m}}\frac{f(\epsilon)}{\epsilon}
\label{eq:Sveps}
\end{equation}
where the pre-factor of the general function $f$ labels the part where the derivative acts on the measure. In a similar fashion to section \ref{ssec:UX} it is useful to extract the measure contribution by
\begin{equation}
\frac{{l+ L+ 2 m}}{{\epsilon+ l+ L+ 2 m}}\frac{f(\epsilon)}{\epsilon}=\left(1-\frac{\epsilon}{{l+L+2m}}+O\left(\epsilon^2\right) \right)\frac{f(\epsilon)}{\epsilon}=\frac{f(\epsilon)}{\epsilon}-\frac{1}{{l+L+2m}}f(\epsilon=0)
\label{eq:separatVX}
\end{equation}
and split \eqref{eq:Sveps}
\begin{equation}
S^v_{\epsilon}=S^v_{\epsilon,\,\text{main}}+S^v_{\epsilon, \text{mes}}
\end{equation}
where the latter part was already evaluated in \cite{usFivePoints}. Concerning the order of summations, it turns out that either $l,\,L,\,k,\,K,\,m$ are valid options. In what follows, we choose the latter possibility which yields more compact expressions. We find the common denominator
\begin{equation}
\begin{split}
d_{S^v_{\epsilon}}=&(1 - a \,\mathbf{q} - b\, \mathbf{s}) (1 - a\, \mathbf{q} - y +  y\,\mathbf{q}) -  \mathbf{s} \,\mathbf{u} \, (a\, b \, y\,\mathbf{q} - a\, b \, y \, \mathbf{q}^2 + z - 2 \, a\,  z\,\mathbf{q} + a^2 \, z \, \mathbf{q}^2 -   b \,z \, \mathbf{s} + a\, b \,    z\,\mathbf{q}\,\mathbf{s}) 
 \end{split}
  \label{eq:DenVX}
\end{equation}
in the four rescalings
\begin{subequations}
\begin{align}
s_{\mathbf{r}}&=\{ a\rightarrow a\,\mathbf{r}, b\rightarrow b, y\rightarrow y\mathbf\,{r}, z\rightarrow z\}, \\ 
s_{\mathbf{r}^2}&=\{ a\rightarrow a\,\mathbf{r}, b\rightarrow b\,\mathbf{r}, y\rightarrow y\,\mathbf{r}, z\rightarrow z\,\mathbf{r}\}, \\
s_{\mathbf{r},\mathbf{t}}&=\{ a\rightarrow a\,\mathbf{r}, b\rightarrow b\,\mathbf{t}, y\rightarrow y\,\mathbf{r}, z\rightarrow z\,\mathbf{t}\},\\ 
s_{\mathbf{r}^2,\mathbf{t}}&=\{ a\rightarrow a\,\mathbf{r}, b\rightarrow b\mathbf{r}\,\mathbf{t}, y\rightarrow y\,\mathbf{r}, z\rightarrow z\,\mathbf{r}\,\mathbf{t}\}.
\end{align}
\end{subequations}

\begin{table}[h]
\begin{center}
\begin{tabular}{ c|c|c|c|c } 

 $ $ & 	$s_{\mathbf{r}}$ & $s_{\mathbf{r}^2}$& $s_{\mathbf{r},\mathbf{t}}$  & $s_{\mathbf{r}^2,\mathbf{t}}$  \\ 
  \hline
   \hline
 $\mathcal{Y}_{22}$ & $\star$ & $ $ &  $ $ & $ $ \\ 
 $\mathcal{Y}_{33}$ & $ \star $ & $\star$&  $ $ &  $ $  \\ 
 $\mathcal{Y}_{44}$ & $ \star $ & $\star$& $ $ & $ $   \\ 
 $\mathcal{Z}_{22}$ & $ $  & $ $ & $\star$ & $\star$   \\ 
 $\mathcal{Z}_{44}$ & $ $  & $ $& $\star$ & $\star$ \\
 \hline
\end{tabular}
\end{center}
\caption{Rescaling rules for the cross-ratios for the $S^v_{\epsilon}$-residues}
\label{tab:rescalingsVX}
\end{table}

Again, from the summation over $l$ we obtain integrations over $\mathbf{q}$ and $\mathbf{r}$ which are present for every matrix element, and the two possible scalings for $\mathbf{r}$ are the result of partially fractioning the rational factors including $l$ into terms with either of the two distinct denominators ${(\epsilon+ l+ L +2 \, m)} $ or ${(\epsilon+ k+ K+ l+ L+ 3 \, m)}$ prior to summation. For the first case we translate $_3F_2$ into a double integral using \eqref{eq:rule3F2a1b1} and \eqref{eq:rule2F1II}, while for the second one \eqref{eq:rule3F2a1b1} and \eqref{eq:rule2F1I} are employed. Next is the summation over $k$ which results in either a $_3F_2$ or a $_2F_1$. The $_3F_2$ is rewritten as a $\ft$ integral over a $_2F_1$ using \eqref{eq:rule3F2a3b2}. Furthermore, we obtain an integration over $\mathbf{s}$ for every contribution as a result of using \eqref{eq:rule2F1I} on the $_2F_1$. In \eqref{eq:DenVX}, an integration over $\mathbf{u}$ is present for all contributions. It is obtained by applying \eqref{eq:rule2F1I} on the $_2F_1$-function from the hypergeometric series over $K$. Finally, the successive series for $L$ and $m$ are always of geometric type.

\subsubsection*{Integration}

The specific structure of the Euler integrals does come with a caveat and is not as straightforward as in the two previously discussed cases. We are not able to simply expand in $\epsilon$ up to the desired order, because the integral over the parameter $\mathbf{s}$ is actually divergent for $\epsilon \rar 0$ and $\fs \eqsp 1$ and appears always in the form
\begin{equation}
I_{s}=\int_0^1(1-\mathbf{s})^{\epsilon-1}f(\mathbf{s},\epsilon).
\end{equation}
We should immediately deal with the divergent integral expanding $f(\mathbf{s},\epsilon)$  up to order $O(\epsilon)$, which can indeed safely been done as $f$ is regular in $\mathbf{s}$. Doing so we may obtain logarithms as $\epsilon$ will most likely be present as an exponent. Followed by a partial fraction decomposition in $\mathbf{s}$ which isolates the $(1-\mathbf{s})$ denominator, the integrations are always 
\begin{subequations}
\begin{equation}
\int_{0}^{1}\frac{1}{1-\mathbf{s}}(1-\mathbf{s})^{\pm \epsilon}\text{d}\mathbf{s}=\pm\frac{1}{\epsilon}
\label{eq:DivIntI}
\end{equation}
(by analytic continuation) or
\begin{equation}
\int_{0}^{1}\frac{1}{1-\mathbf{s}}(1-\mathbf{s})^{\pm \epsilon} \, \log[a+b*\mathbf{s}]\text{d}\mathbf{s}=\pm\frac{\log[a + b]}{\epsilon} +O(\epsilon^0).
\label{eq:DivIntII}
\end{equation}
\end{subequations}
The $O(\epsilon^0)$ terms in \eqref{eq:DivIntII} will lead to hypergeometric functions, but luckily we only need to work up to $O(\epsilon^{-1})$. The reason being that the logarithms in $f$ contribute at $O(\epsilon^1)$, so only in combination with the $O(\epsilon^{-1})$-term in \eqref{eq:DivIntII}, a minimal order $O(\epsilon^0)$ is obtained. Higher orders in $\epsilon$ can safely be omitted since $\epsilon \to 0$ and it is safe to continue with the next integrals. From that point onward the integration process is completely regular. Once the integration over $\mathbf{s}$ has been evaluated, the structure of the defining denominator implies an integration over $\mathbf{u}$. After that the remaining integrations are simple.

\subsection{Re-summation of $S(\mathcal{M})^v_{\psi}$}

The steps presented in Section  \ref{hardest} can be applied in the very same manner to the contributions of $\mathcal{Y}_{11}$, $\mathcal{Y}_{33}$, $\mathcal{Y}_{44}$ and $\mathcal{Z}_{44}$.  While the complexity of the $\mathcal{Z}_{44}$ case matches for $\mathcal{Z}_{33}$, the three $\mathcal{Y}$-elements are far simpler to handle, both in terms of summation and subsequent integration. All Euler integrals contain the denominator $P_1$ found in section \eqref{eq:P1} with :
\begin{equation}
d_{S^v_{\psi}}=P_1   \label{eq:LongDen}
 \end{equation}
 with the rescaling options:
\begin{subequations}
\begin{align}
s_{0}&=\{a\rightarrow a, b\rightarrow b, y\rightarrow y, z\rightarrow z\},\\
s_{\mathbf{r}^3}&=\{a\rightarrow a\,\mathbf{q}, b\rightarrow b\,\mathbf{r}, y\rightarrow y\,\mathbf{r}, z\rightarrow z\,\mathbf{r}\},\\
s_{\mathbf{r}^2}&=\{a\rightarrow a\,\mathbf{r}, b\rightarrow b, y\rightarrow y\,\mathbf{r}, z\rightarrow z\},\\
s_{\mathbf{t}^2}&=\{a\rightarrow a, b\rightarrow b\,\mathbf{t}, y\rightarrow y, z\rightarrow z\,\mathbf{t}\},\\
s_{\mathbf{r}^2,\,\mathbf{t}^2}&=\{a\rightarrow a\,\mathbf{r}, b\rightarrow b\,\mathbf{t}, y\rightarrow y\,\mathbf{q}, z\rightarrow z\,\mathbf{t}\},\\
s_{\mathbf{r}^3,\,\mathbf{t}^2}&=\{a\rightarrow a\,\mathbf{r}, b\rightarrow b\,\mathbf{r}\,\mathbf{t}, y\rightarrow y\,\mathbf{r}, z\rightarrow z\,\mathbf{r}\,\mathbf{t}\}.
\end{align}
\end{subequations}
The subscript indicates whether the Euler parameter appears quadratic or cubic after $P_1$ is sent to its original form. We also send $a_0\rightarrow a$ for all transformations displayed above. Their relevance to the remaining $S^v_{\psi}$ is listed in Table \ref{tab:EfDenForPoliV}.
\begin{table}[h]
\begin{center}
\begin{tabular}{ c|c|c|c|c|c|c } 
 $ $ & 	$s_{0}$ & $s_{\mathbf{t}^2}$& $s_{\mathbf{r}^2}$ & $s_{\mathbf{r}^3}$  & $s_{\mathbf{r}^2,\,\mathbf{t}^2}$ & $s_{\mathbf{r}^3,\,\mathbf{t}^2}$ \\ 
  \hline
   \hline
 $\mathcal{Y}_{11}$ & $ $ & $\star$ & $ \star$ & $ $&  $ $ & $ $   \\ 
 $\mathcal{Y}_{33}$ & $\star$ & $ $&  $\star$ &  $\star$& $\star$ & $\star$ \\ 
 $\mathcal{Y}_{44}$ & $ $ & $ $& $\star$ & $\star$& $ $ & $ $ \\ 
 $\mathcal{Z}_{44}$ & $ $ & $ $& $\star$ & $\star$& $\star$ & $\star$  \\
 \hline
\end{tabular}
\end{center}
\caption{Rescaling rules for the cross-ratios for the $S^v_{\psi}$-residues}
\label{tab:EfDenForPoliV}
\end{table}

\section{Function space}

The function space which was derived in \cite{usFivePoints} failed to give solutions to the $S_{\psi}$-residues. As a consequence it was only possible to state explicit results for the sum of ${S(\mathcal{Y}_{11})+S(\mathcal{Y}_{33})}$, ${S(\mathcal{Y}_{22})+S(\mathcal{Y}_{44})}$ and 
$S(\mathcal{Z}_{11})+S(\mathcal{Z}_{22})+S(\mathcal{Z}_{33})+S(\mathcal{Z}_{44})$. The methods presented in the previous section enabled us to analytically derive the result for every residue including the $S_{\psi}$. As a consequence, with the explicit forms of $S(\mathcal{Y}_{11})$ to $S(\mathcal{Y}_{44})$ as well as $S(\mathcal{Z}_{22})$ to $S(\mathcal{Z}_{44})$ in hand, we are able to expand and complete function space from \cite{usFivePoints}. 
\subsubsection*{Set of denominators}
The results stated in section \ref{sec:resultsYZ} can be written over denominators which belong to the set:
\begin{equation}
\begin{split}
\{&
 a-y,\,b-z, 
 b\,y-a\,z,\,
 a-y+b \, y-a \, z,\,
 b-b\,y-z+a\,z,\\&
a\,b-a\,b\,y-a\,b\,z-y\,z+a\,y\,z+b\,y\,z,\,\\&
a\,b-a\,b\,y-y\,z+a\,y\,z,\, a\,b-a\,b\,z-y\,z+b\,y\,z \}.
\end{split}
\end{equation}
Although $S_{\psi}$ was far from being accessible with the techniques developed in  \cite{usFivePoints}, it turns out that there is no additional denominator which would cancel in the sum $S({\mathcal{Y}_{11}}) +S({\mathcal{Y}_{33}})$,  $S({\mathcal{Y}_{22}}) +S({\mathcal{Y}_{44}})$ and  $S({\mathcal{Z}_{11}}) +S({\mathcal{Z}_{22}})+S({\mathcal{Z}_{33}})+S({\mathcal{Z}_{44}})$. Therefore, the set of denominators derived in  \cite{usFivePoints} was already complete.
 
\subsubsection*{Symbol letters}
With every contribution $S(\mathcal{M})$ being fully available now, we can complete the symbol alphabet stated in \cite{usFivePoints}. For the first letter we find the following twelve entries
\begin{equation}
\begin{split}
l1\in\{&a,\,b,\,y,\,z,\,1-a,\,1-b,\,1-y,\,1-z,\,1-a-b,\,1-y-z,\\
&a-y,\,b-z,\,a - y + b\, y - a\, z,\,b - b\, y - z + a \,z,\\
&a \,b - 2 \,a\, b \,y + a \,b \,y^2 - a\, b \,z - y \,z + 2\, a\, y \,z - a^2 \,y \,z + b \,y\, z,\\
&a \,b - a \,b\, y - 2\, a\, b\, z - y\, z + a\, y \,z + 2\, b\, y \,z - b^2 \,y \,z + a\, b\, z^2,\\
&a\,  b - a\,  b \, y - a  \,b \, z - y \, z + a \, y \, z + b  \,y  \,z\}.
\end{split}
\end{equation}
Here, $a \, b - 2 \, a \, b \, y + a \, b \, y^2 - a \, b \, z - y \, z + 2 \, a \, y \, z - a^2 \, y \, z + b \, y \, z$ and $a \, b - a \, b \, y - 2 \, a \, b \, z - y \, z + a \, y \, z + 2 \, b \, y \, z - b^2 \, y \, z + a \, b \, z^2$ are only obtained in the $S_{\psi}$ calculations and are the only new first letters of the expanded alphabet.

The second letters for the symbols either take the form of $1-\dots$ or contain solely the four cross-ratios. The latter is a consequence of the derivative acting on the weight factor which instantly results in a logarithm of said cross-ratios. We find the following complete set
\begin{equation}
\begin{split}
l2\in\{&a,\,b,\,y,\,z,\,1-a,\,1-b,\,1-y,\,1-z,\,1-a-b,\,1-y-z,
\\&1-b-a \, y, \,1-a \, y-z,\,1-a-b \, z,\,1-y-b \, z\}.
\end{split}
\end{equation}
The new entries which are exclusive to $S_{\psi}$, are the last four in the list and always contain a product of two cross-ratios. Additionally, unlike the remaining 10 letters, they do not appear in the first symbol entry.

\section{Explicit results for the re-summation}
\label{sec:resultsYZ}
Using the methods presented in the previous sections, we can finally derive analytical results for the previously unknown contribution $S(\mathcal{Y}_{11})$ to $S(\mathcal{Y}_{44})$ and $S(\mathcal{Z}_{22})$ to $S(\mathcal{Z}_{44})$. In the following we will present the explicit results. Instead of showing the complicated and lengthy expression in terms of dilogarithms, we rely on their symbol once again. In order to make the expressions more pleasant to read, we introduce some notation for the symbol entries:
\begin{equation}
l_{x\,\dots\,(yx)} = 1-x-\,\dots\,-y\,z
\end{equation}
and define one last polynomial
\begin{equation}
D_{14}=a \,b - a\, b\, y - 2\, a\, b\, z - y\, z + a \,y \,z + 2\, b\, y\, z - b^2\, y \,z + a \,b\, z^2.
\end{equation}
\subsubsection*{Explicit form of $S(\mathcal{Y}_{11})$}

\begin{subequations}
\begin{equation}
\mathcal{S}_{\mathcal{Y}_{11}}= 
-g^2\frac{z}{b-z}\mathcal{S}_{\mathcal{Y}_{11}}^{\text{I}}+
g^2\frac{z}{b-b y-z+a z}\mathcal{S}_{\mathcal{Y}_{11}}^{\text{II}}+
g^2\frac{a y z}{a b-a b y-a b z-y z+a y z+b y z}\mathcal{S}_{\mathcal{Y}_{11}}^{\text{III}}
\end{equation}
\begin{align}
\mathcal{S}_{\mathcal{Y}_{11}}^{\text{I}}&=
-\mathcal{S}\left[b\otimes \left(l_b l_z\right)\right]+\mathcal{S}\left[z\otimes \left(l_b l_z\right)\right]+\mathcal{S}\left[l_b\otimes (b z)\right]-\mathcal{S}\left[l_z\otimes (b z)\right]
\\&
&\nonumber\\
\mathcal{S}_{\mathcal{Y}_{11}}^{\text{II}}&=
\mathcal{S}\left[b, \frac{l_y}{l_b l_{y z}}\right]+\mathcal{S}\left[(a-y), \frac{l_y}{l_a}\right]+\mathcal{S}\left[(b-z), \frac{l_b l_{(a y) z}}{l_{b (a y)}
   l_z}\right]+\mathcal{S}\left[z, \frac{l_{a b} l_z}{l_a}\right]+\mathcal{S}\left[l_a, \frac{l_{a b} l_{(a y) z}}{b z}\right]\\&+\mathcal{S}\left[l_{a b}, \frac{b z l_b}{l_a
   l_{b (a y)}}\right]+\mathcal{S}\left[l_y, \frac{b z}{l_{b (a y)} l_{y z}}\right]+\mathcal{S}\left[l_{y z}, \frac{l_y l_{(a y) z}}{b z
   l_z}\right]+\mathcal{S}\left[D_1, \frac{l_{a b} l_z}{l_b l_{y z}}\right]+\left[D_5, \frac{l_a l_{b (a y)} l_{y z}}{l_{a b} l_y l_{(a y) z}}\right]\nonumber\\
&\nonumber\\
\mathcal{S}_{\mathcal{Y}_{11}}^{\text{III}}&=
\mathcal{S}\left[a, \frac{l_a l_{b (a y)} l_z}{l_{a b} l_{(a y) z}}\right]+\mathcal{S}\left[b, \frac{l_a l_b}{l_{a
   b}}\right]+\mathcal{S}\left[(a-y), \frac{l_a}{l_y}\right]+\mathcal{S}\left[y, \frac{l_{b (a y)} l_{y z}}{l_b l_y l_{(a y) z}}\right]
   +\mathcal{S}\left[(b-z), \frac{l_{b   (a y)} l_z}{l_b l_{(a y) z}}\right]
   \\&+\mathcal{S}\left[z, \frac{l_{y z}}{l_y l_z}\right]+\mathcal{S}\left[l_a, \frac{l_{y z}}{l_{(a y)
   z}}\right]+\mathcal{S}\left[l_b, \frac{l_{a b}}{l_{b (a y)}}\right]+\mathcal{S}\left[l_{a b}, \frac{l_{b (a y)}}{l_b l_y}\right]+\mathcal{S}\left[l_y, \frac{l_{b (a
   y)}}{l_{a b}}\right]\nonumber\\&+\mathcal{S}\left[l_z, \frac{l_{(a y) z}}{l_{y z}}\right]+\mathcal{S}\left[l_{y z}, \frac{l_a l_z}{l_{(a y)
   z}}\right]+\mathcal{S}\left[D_1, \frac{l_b l_{y z}}{l_{a b} l_z}\right]+\mathcal{S}\left[D_5, \frac{l_{a b} l_y l_{(a y) z}}{l_a l_{b (a y)} l_{y z}}\right]\nonumber
\end{align}
\end{subequations}

\subsubsection*{Explicit form of $S(\mathcal{Y}_{22})$}
\begin{subequations}
\begin{equation}
\mathcal{S}_{\mathcal{Y}_{22}}= 
-g^2\frac{y}{a-y}\mathcal{S}_{\mathcal{Z}_{22}}^{\text{I}}+
g^2\frac{y}{a-y+b y-a z}\mathcal{S}_{\mathcal{Z}_{22}}^{\text{II}}+
g^2\frac{b y z}{a b-a b y-a b z-y z+a y z+b y z}\mathcal{S}_{\mathcal{Z}_{22}}^{\text{III}}
\end{equation}
\begin{align}
\mathcal{S}_{\mathcal{Y}_{22}}^{\text{I}}&=
-\mathcal{S}\left[a, \left(l_a l_y\right)\right]+\mathcal{S}\left[y, \left(l_a l_y\right)\right]+\mathcal{S}\left[l_a, (a y)\right]-\mathcal{S}\left[l_y, (a y)\right]
\\&
&\nonumber\\
\mathcal{S}_{\mathcal{Y}_{22}}^{\text{II}}&=
-\mathcal{S}\left[a, \frac{l_a l_{y z}}{l_z}\right]-\mathcal{S}\left[(a-y), \frac{l_y l_{a (b z)}}{l_a l_{y (b z)}}\right]-\mathcal{S}\left[y, \frac{l_b}{l_{a b}
   l_y}\right]-\mathcal{S}\left[(b-z), \frac{l_b}{l_z}\right]-\mathcal{S}\left[l_b, \frac{a y}{l_{a b} l_{y (b z)}}\right]\\&-\mathcal{S}\left[l_{a b}, \frac{l_b l_{a (b
   z)}}{a y l_a}\right]-\mathcal{S}\left[l_z, \frac{l_{y z} l_{a (b z)}}{a y}\right]-\mathcal{S}\left[l_{y z}, \frac{a y l_y}{l_z l_{y (b
   z)}}\right]-\mathcal{S}\left[D_2, \frac{l_a l_{y z}}{l_{a b} l_y}\right]-\mathcal{S}\left[D_{14}, \frac{l_{a b} l_z l_{y (b z)}}{l_b l_{y z} l_{a (b z)}}\right]\nonumber\\
&\nonumber\\
\mathcal{S}_{\mathcal{Y}_{22}}^{\text{III}}&=
\mathcal{S}\left[a, \frac{l_a l_b}{l_{a b}}\right]+\mathcal{S}\left[b, \frac{l_b l_y l_{a (b z)}}{l_{a b} l_{y (b z)}}\right]+\mathcal{S}\left[(a-y), \frac{l_y l_{a
   (b z)}}{l_a l_{y (b z)}}\right]+\mathcal{S}\left[y, \frac{l_{y z}}{l_y l_z}\right]+\mathcal{S}\left[(b-z), \frac{l_b}{l_z}\right]\\&+\mathcal{S}\left[z, \frac{l_{y z} l_{a
   (b z)}}{l_a l_z l_{y (b z)}}\right]+\mathcal{S}\left[l_a, \frac{l_{a b}}{l_{a (b z)}}\right]+\mathcal{S}\left[l_b, \frac{l_{y z}}{l_{y (b z)}}\right]+\mathcal{S}\left[l_{a
   b}, \frac{l_{a (b z)}}{l_a l_z}\right]+\mathcal{S}\left[l_y, \frac{l_{y (b z)}}{l_{y z}}\right]\\&+\mathcal{S}\left[l_z, \frac{l_{a (b z)}}{l_{a
   b}}\right]+\mathcal{S}\left[l_{y z}, \frac{l_b l_y}{l_{y (b z)}}\right]+\mathcal{S}\left[D_2, \frac{l_a l_{y z}}{l_{a b} l_y}\right]+\mathcal{S}\left[D_{14}, \frac{l_{a b}
   l_z l_{y (b z)}}{l_b l_{y z} l_{a (b z)}}\right]
&\nonumber\\
\end{align}
\end{subequations}

\subsubsection*{Explicit form of $S(\mathcal{Y}_{33})$}

\begin{subequations}
\begin{equation}
\mathcal{S}_{\mathcal{Y}_{33}}= 
g^2\frac{z}{b-b y-z+a z}\mathcal{S}_{\mathcal{Y}_{33}}^{\text{I}}+
-g^2\frac{a y z}{a b-a b y-a b z-y z+a y z+b y z}\mathcal{S}_{\mathcal{Y}_{33}}^{\text{II}}
\end{equation}

\begin{align}
\mathcal{S}_{\mathcal{Y}_{33}}^{\text{I}}&=\mathcal{S}\left[b, \frac{l_a l_b}{l_{a b}}\right]+\mathcal{S}\left[(a-y), \frac{l_a}{l_y}\right]+\mathcal{S}\left[(b-z), \frac{l_{b (a y)} l_z}{l_b l_{(a y)
   z}}\right]+\mathcal{S}\left[z, \frac{l_{y z}}{l_y l_z}\right]+\mathcal{S}\left[l_a, \frac{l_{y z}}{l_{(a y) z}}\right]\\&+\mathcal{S}\left[l_{a b}, \frac{l_{b (a y)}}{l_b
   l_y}\right]+\mathcal{S}\left[l_y, \frac{l_{b (a y)}}{l_{a b}}\right]+\mathcal{S}\left[l_{y z}, \frac{l_a l_z}{l_{(a y) z}}\right]+\mathcal{S}\left[D_1, \frac{l_b l_{y
   z}}{l_{a b} l_z}\right]+\mathcal{S}\left[D_5, \frac{l_{a b} l_y l_{(a y) z}}{l_a l_{b (a y)} l_{y z}}\right]\nonumber
 \end{align}
\begin{align}
\mathcal{S}_{\mathcal{Y}_{33}}^{\text{II}}&=\mathcal{S}\left[a, \frac{l_{b (a y)} l_{y z}}{l_b l_y l_{(a y) z}}\right]+\mathcal{S}\left[b, \frac{l_{y z}}{l_y
   l_z}\right]+\mathcal{S}\left[(a-y), \frac{l_a}{l_y}\right]+\mathcal{S}\left[y, \frac{l_a l_{b (a y)} l_z}{l_{a b} l_{(a y) z}}\right]+\mathcal{S}\left[(b-z), \frac{l_{b
   (a y)} l_z}{l_b l_{(a y) z}}\right]\\&+\mathcal{S}\left[z, \frac{l_a l_b}{l_{a b}}\right]+\mathcal{S}\left[l_a, \frac{a b y z}{l_{a b} l_{(a y)
   z}}\right]+\mathcal{S}\left[l_b, \frac{a b y z}{l_{b (a y)} l_{y z}}\right]+\mathcal{S}\left[l_{a b}, \frac{l_a l_{b (a y)} l_z}{a b y
   z}\right]+\mathcal{S}\left[l_y, \frac{l_{b (a y)} l_{y z}}{a b y z}\right]\nonumber\\&+\mathcal{S}\left[l_z, \frac{l_{a b} l_{(a y) z}}{a b y z}\right]+\mathcal{S}\left[l_{y
   z}, \frac{a b y z}{l_b l_y l_{(a y) z}}\right]+\mathcal{S}\left[D_1, \frac{l_b l_{y z}}{l_{a b} l_z}\right]+\mathcal{S}\left[D_5, \frac{l_{a b}
   l_y l_{(a y) z}}{l_a l_{b (a y)} l_{y z}}\right]\nonumber
\end{align}
\end{subequations}

\subsubsection*{Explicit form of $S(\mathcal{Y}_{44})$}

\begin{subequations}
\begin{equation}
\mathcal{S}_{\mathcal{Y}_{44}}= 
g^2\frac{y}{a-y+b y-a z}\mathcal{S}_{\mathcal{Y}_{44}}^{\text{I}}+
-g^2\frac{b y z}{a b-a b y-a b z-y z+a y z+b y z}\mathcal{S}_{\mathcal{Y}_{44}}^{\text{II}}
\end{equation}
\begin{align}
\mathcal{S}_{\mathcal{Y}_{44}}^{\text{I}}&=\mathcal{S}\left[a, \frac{l_a l_b}{l_{a b}}\right]+\mathcal{S}\left[(a-y), \frac{l_y l_{a (b z)}}{l_a l_{y (b z)}}\right]+
\mathcal{S}\left[y, \frac{l_{y z}}{l_y   l_z}\right]+\mathcal{S}\left[(b-z), \frac{l_b}{l_z}\right]+\mathcal{S}\left[l_b, \frac{l_{y z}}{l_{y (b z)}}\right]\\&+\mathcal{S}\left[l_{a b}, \frac{l_{a (b z)}}{l_a   l_z}\right]+\mathcal{S}\left[l_z, \frac{l_{a (b z)}}{l_{a b}}\right]+\mathcal{S}\left[l_{y z}, \frac{l_b l_y}{l_{y (b z)}}\right]+\mathcal{S}\left[D_2, \frac{l_a l_{y   z}}{l_{a b} l_y}\right]+\mathcal{S}\left[D_{14}, \frac{l_{a b} l_z l_{y (b z)}}{l_b l_{y z} l_{a (b z)}}\right]\nonumber
\\&
&\nonumber\\
\mathcal{S}_{\mathcal{Y}_{44}}^{\text{II}}&=\mathcal{S}\left[a, \frac{l_{y z}}{l_y l_z}\right]+\mathcal{S}\left[b, \frac{l_{y z} l_{a (b z)}}{l_a l_z l_{y (b z)}}\right]+\mathcal{S}\left[(a-y), \frac{l_y l_{a  (b z)}}{l_a l_{y (b z)}}\right]+\mathcal{S}\left[y, \frac{l_a l_b}{l_{a b}}\right]+\mathcal{S}\left[(b-z), \frac{l_b}{l_z}\right]\\&+\mathcal{S}\left[z, \frac{l_b l_y l_{a
   (b z)}}{l_{a b} l_{y (b z)}}\right]+\mathcal{S}\left[l_a, \frac{a b y z}{l_{y z} l_{a (b z)}}\right]+\mathcal{S}\left[l_b, \frac{a b y z}{l_{a b} l_{y (b
   z)}}\right]+\mathcal{S}\left[l_{a b}, \frac{l_b l_y l_{a (b z)}}{a b y z}\right]+\mathcal{S}\left[l_y, \frac{l_{a b} l_{y (b z)}}{a b y
   z}\right]\nonumber\\&+\mathcal{S}\left[l_z, \frac{l_{y z} l_{a (b z)}}{a b y z}\right]+\mathcal{S}\left[l_{y z}, \frac{a b y z}{l_a l_z l_{y (b
   z)}}\right]+\mathcal{S}\left[D_2, \frac{l_a l_{y z}}{l_{a b} l_y}\right]+\mathcal{S}\left[D_{14}, \frac{l_{a b} l_z l_{y (b z)}}{l_b l_{y z} l_{a (b z)}}\right]\nonumber
\end{align}
\end{subequations}

\subsubsection*{Explicit form of $S(\mathcal{Z}_{22})$}

\begin{subequations}
\begin{equation}
\setlength{\jot}{8pt}
\begin{gathered}
\mathcal{S}_{\mathcal{Z}_{22}}= 
g^2\mathcal{S}_{\mathcal{Z}_{22}}^{\text{I}}+
g^2\frac{-a}{a-y}\mathcal{S}_{\mathcal{Z}_{22}}^{\text{II}}-
g^2\frac{a (1-z)}{a-y+b y-a z}\mathcal{S}_{\mathcal{Z}_{22}}^{\text{III}}-
g^2\frac{2 a z}{b y-a z}\mathcal{S}_{\mathcal{Z}_{22}}^{\text{IV}}\\
+g^2\frac{a b (1-y-z)}{-a b+a b y+a b z+y z-a y z-b y z}\mathcal{S}_{\mathcal{Z}_{22}}^{\text{V}}
\end{gathered}
\end{equation}

\begin{align}
\mathcal{S}_{\mathcal{Z}_{22}}^{\text{I}}&=\mathcal{S}\left[b\otimes \frac{l_a^2 l_b^2}{l_{a b}^2}\right]+\mathcal{S}\left[(a-y), \frac{l_{a (b z)}}{l_{y (b z)}}\right]+\mathcal{S}\left[y, \frac{l_a l_b}{l_{a
   b}}\right]+\mathcal{S}\left[z, \frac{l_{y z} l_{a (b z)}}{l_a l_z l_{y (b z)}}\right]+\mathcal{S}\left[l_a, \frac{a y l_{a b}}{l_{a (b
   z)}}\right]\\&+\mathcal{S}\left[l_b, \frac{a y l_{y z}}{l_b l_{y (b z)}^2}\right]+\mathcal{S}\left[l_{a b}, \frac{l_b l_{a (b z)}^2}{a y l_a^2
   l_z}\right]+\mathcal{S}\left[D_1, \frac{l_b l_{y z}}{l_{a b} l_z}\right]+\mathcal{S}\left[D_{14}, \frac{l_{a b} l_z l_{y (b z)}}{l_b l_{y z} l_{a (b z)}}\right]\nonumber\\
&\nonumber\\
\mathcal{S}_{\mathcal{Z}_{22}}^{\text{II}}&=-\mathcal{S}\left[a, l_y\right]+\mathcal{S}\left[y, l_a\right]+\mathcal{S}\left[l_a, (a y)\right]-\mathcal{S}\left[l_y, (a y)\right]\\
      &\nonumber\\
\mathcal{S}_{\mathcal{Z}_{22}}^{\text{III}}&=\mathcal{S}\left[a, \frac{l_a l_b l_{y z}}{l_{a b} l_z}\right]+\mathcal{S}\left[(a-y), \frac{l_y l_{a (b z)}}{l_a l_{y (b z)}}\right]+\mathcal{S}\left[y, \frac{l_b
   l_{y z}}{l_{a b} l_y l_z}\right]+\mathcal{S}\left[(b-z), \frac{l_b}{l_z}\right]+\mathcal{S}\left[l_b, \frac{a y}{l_b l_{y (b z)}}\right]\\&+\mathcal{S}\left[l_{a
   b}, \frac{l_b l_{a (b z)}}{a y l_a}\right]+\mathcal{S}\left[l_z, \frac{l_z l_{a (b z)}}{a y}\right]+\mathcal{S}\left[l_{y z}, \frac{a y l_y}{l_z
   l_{y (b z)}}\right]+\mathcal{S}\left[D_2, \frac{l_a l_{y z}}{l_{a b} l_y}\right]+\mathcal{S}\left[D_{14}, \frac{l_{a b} l_z l_{y (b z)}}{l_b l_{y z} l_{a (b
   z)}}\right]\nonumber\\
&\nonumber\\
\mathcal{S}_{\mathcal{Z}_{22}}^{\text{IV}}&=\mathcal{S}\left[a, \frac{l_a l_b}{l_{a b}}\right]+\mathcal{S}\left[b, \frac{l_{a b}}{l_a
   l_b}\right]+\mathcal{S}\left[(a-y), \frac{l_y}{l_a}\right]+\mathcal{S}\left[y, \frac{l_{y z}}{l_y
   l_z}\right]\\&+\mathcal{S}\left[(b-z), \frac{l_b}{l_z}\right]+\mathcal{S}\left[z, \frac{l_y l_z}{l_{y z}}\right]+\mathcal{S}\left[D_1, \frac{l_{a b} l_z}{l_b l_{y
   z}}\right]+\mathcal{S}\left[D_2, \frac{l_a l_{y z}}{l_{a b} l_y}\right]\nonumber
\end{align}

\begin{align}   
      \mathcal{S}_{\mathcal{Z}_{22}}^{\text{V}}&=\mathcal{S}\left[a, \frac{l_a l_b}{l_{a b}}\right]+\mathcal{S}\left[b, \frac{l_b l_y l_{a (b z)}}{l_{a b} l_{y (b z)}}\right]+\mathcal{S}\left[(a-y), \frac{l_y l_{a   (b z)}}{l_a l_{y (b z)}}\right]+\mathcal{S}\left[y, \frac{l_{y z}}{l_y l_z}\right]+\mathcal{S}\left[(b-z), \frac{l_b}{l_z}\right]\\&+\mathcal{S}\left[z, \frac{l_{y z} l_{a
   (b z)}}{l_a l_z l_{y (b z)}}\right]+\mathcal{S}\left[l_a, \frac{l_{a b}}{l_{a (b z)}}\right]+\mathcal{S}\left[l_b, \frac{l_{y z}}{l_{y (b z)}}\right]+\mathcal{S}\left[l_{a
   b}, \frac{l_{a (b z)}}{l_a l_z}\right]+\mathcal{S}\left[l_y, \frac{l_{y (b z)}}{l_{y z}}\right] \nonumber \\&+\mathcal{S}\left[l_z, \frac{l_{a (b z)}}{l_{a
   b}}\right]+\mathcal{S}\left[l_{y z}, \frac{l_b l_y}{l_{y (b z)}}\right]+\mathcal{S}\left[D_2, \frac{l_a l_{y z}}{l_{a b} l_y}\right]+\mathcal{S}\left[D_{14}, \frac{l_{a b}
   l_z l_{y (b z)}}{l_b l_{y z} l_{a (b z)}}\right]\nonumber
\end{align}
\end{subequations}

\subsubsection*{Explicit form of $S(\mathcal{Z}_{33})$}

\begin{equation}
\setlength{\jot}{8pt}
\begin{gathered}
\mathcal{S}_{\mathcal{Z}_{33}}= 
g^2\mathcal{S}_{\mathcal{Z}_{33}}^{\text{I}}+
g^2\frac{-b}{b-z}\mathcal{S}_{\mathcal{Z}_{33}}^{\text{II}}+
g^2\frac{-2 b y}{b y-a z}\mathcal{S}_{\mathcal{Z}_{33}}^{\text{III}}+
g^2\frac{b (1-y)}{a z-b y+b-z}\mathcal{S}_{\mathcal{Z}_{33}}^{\text{IV}}\\+
g^2\frac{a b (1-y-z)}{a b y+a b z-a b-a y z-b y z+y z}\mathcal{S}_{\mathcal{Z}_{33}}^{\text{V}}
\end{gathered}
\end{equation}

\begin{subequations}
\begin{align}
\mathcal{S}_{\mathcal{Z}_{33}}^{\text{I}}&=\mathcal{S}\left[a, \frac{l_a^2 l_b^2}{l_{a b}^2}\right]+\mathcal{S}\left[y, \frac{l_{b (a  y)} l_{y z}}{l_b l_y l_{(a y) z}}\right]+\mathcal{S}\left[(b-z), \frac{l_{b (a   y)}}{l_{(a y) z}}\right]+\mathcal{S}\left[z, \frac{l_a l_b}{l_{a  b}}\right]+\mathcal{S}\left[l_a, \frac{b z l_{y z}}{l_a l_{(a y)   z}^2}\right]\\&
+\mathcal{S}\left[l_b, \frac{b z l_{a b}}{l_{b (a y)}}\right]+\mathcal{S}\left[l_{a   b}, \frac{l_a l_{b (a y)}^2}{b z l_b^2   l_y}\right]+\mathcal{S}\left[D_2, \frac{l_a l_{y z}}{l_{a b}   l_y}\right]
+\mathcal{S}\left[D_5, \frac{l_{a b} l_y l_{(a y) z}}{l_a l_{b (a y)} l_{y   z}}\right]\nonumber\\
&\nonumber\\
\mathcal{S}_{\mathcal{Z}_{33}}^{\text{II}}&=-\mathcal{S}\left[b, l_z\right]+\mathcal{S}\left[z, l_b\right]+\mathcal{S}\left[l_b, (b  z)\right]-\mathcal{S}\left[l_z\otimes (b z)\right]&\\
&\nonumber\\
\mathcal{S}_{\mathcal{Z}_{33}}^{\text{III}}&=\mathcal{S}\left[a, \frac{l_a l_b}{l_{a b}}\right]+\mathcal{S}\left[b, \frac{l_{a b}}{l_a
   l_b}\right]+\mathcal{S}\left[(a-y), \frac{l_y}{l_a}\right]+\mathcal{S}\left[y, \frac{l_{y z}}{l_y
   l_z}\right]\\&
   +\mathcal{S}\left[(b-z), \frac{l_b}{l_z}\right]+\mathcal{S}\left[z, \frac{l_yl_z}{l_{y z}}\right]+\mathcal{S}\left[D_1, \frac{l_{a b} l_z}{l_b l_{y   z}}\right]+\mathcal{S}\left[D_2, \frac{l_a l_{y z}}{l_{a b} l_y}\right]\nonumber\\
   &\nonumber\\
\mathcal{S}_{\mathcal{Z}_{33}}^{\text{IV}}&=\mathcal{S}\left[b, \frac{l_{a b} l_y}{l_a l_b l_{y
   z}}\right]+\mathcal{S}\left[(a-y), \frac{l_y}{l_a}\right]+\mathcal{S}\left[(b-z), \frac{l_b l_{(a
   y) z}}{l_{b (a y)} l_z}\right]+\mathcal{S}\left[z, \frac{l_{a b} l_y l_z}{l_a l_{y
   z}}\right]+\mathcal{S}\left[l_a, \frac{l_a l_{(a y) z}}{b z}\right]\\&
   +\mathcal{S}\left[l_{ab}, \frac{b z l_b}{l_a l_{b (a y)}}\right]+\mathcal{S}\left[l_y, \frac{b
   z}{l_y l_{b (a y)}}\right]+\mathcal{S}\left[l_{y z}, \frac{l_y l_{(a y) z}}{b z   l_z}\right]
   +\mathcal{S}\left[D_1, \frac{l_{a b} l_z}{l_b l_{yz}}\right]
   +\mathcal{S}\left[D_5, \frac{l_a l_{b (a y)} l_{y z}}{l_{a b} l_y l_{(a y)z}}\right]\nonumber\\ 
      &\nonumber\\
   \mathcal{S}_{\mathcal{Z}_{33}}^{\text{V}}&=
  \mathcal{S}\left[ a, \frac{l_a l_{b (a y)} l_z}{l_{a b} l_{(a y)  z}}\right]+\mathcal{S}\left[b, \frac{l_a l_b}{l_{a   b}}\right]+\mathcal{S}\left[(a-y), \frac{l_a}{l_y}\right]
   +\mathcal{S}\left[y, \frac{l_{b (a y)}l_{y z}}{l_b l_y l_{(a y) z}}\right]+\mathcal{S}\left[(b-z), \frac{l_{b (a y)}   l_z}{l_b l_{(a y) z}}\right]\\&
   +\mathcal{S}\left[z, \frac{l_{y z}}{l_yl_z}\right]+\mathcal{S}\left[l_a, \frac{l_{y z}}{l_{(a y)   z}}\right]+\mathcal{S}\left[l_b, \frac{l_{a b}}{l_{b (a y)}}\right]
   +\mathcal{S}\left[l_{ab}, \frac{l_{b (a y)}}{l_b l_y}\right]+\mathcal{S}\left[l_y, \frac{l_{b   (a y)}}{l_{a b}}\right]\nonumber\\&
   +\mathcal{S}\left[l_z, \frac{l_{(a y) z}}{l_{y z}}\right]+\mathcal{S}\left[l_{yz}, \frac{l_a l_z}{l_{(a y) z}}\right]+\mathcal{S}\left[D_1, \frac{l_b   l_{y z}}{l_{a b} l_z}\right]
   +\mathcal{S}\left[D_5\otimes \frac{l_{a b} l_y l_{(a y) z}}{l_al_{b (a y)} l_{y z}}\right]\nonumber
   \end{align}
\end{subequations}

\vskip 1 cm

\subsubsection*{Explicit form of $\mathcal{Z}_{44}$}

\begin{equation}
\setlength{\jot}{8pt}
\begin{gathered}
\mathcal{S}_{\mathcal{Z}_{44}}= 
g^2\mathcal{S}_{\mathcal{Z}_{44}}^{\text{I}}+
+g^2\frac{ (b y+a z)}{b y-a z}\mathcal{S}_{\mathcal{Z}_{44}}^{\text{II}}+
g^2\frac{b (1-y)}{b-b y-z+a z}\mathcal{S}_{\mathcal{Z}_{44}}^{\text{III}}+
g^2\frac{a (1-z)}{a-y+b y-a z}\mathcal{S}_{\mathcal{Z}_{44}}^{\text{IV}}\\
+g^2\frac{a b (1-y-z)}{-a b y-a b z+a b+a y z+b y z-y z}\mathcal{S}_{\mathcal{Z}_{44}}^{\text{V}}
\end{gathered}
\end{equation}

\begin{subequations}
\begin{align}
\mathcal{S}_{\mathcal{Z}_{44}}^{\text{I}}&=\mathcal{S}\left[a, \frac{l_a l_b}{l_{a b}}\right]+\mathcal{S}\left[b, \frac{l_a l_b}{l_{a b}}\right]+\mathcal{S}\left[(a-y), \frac{l_{a (b z)}}{l_{y (b
   z)}}\right]+\mathcal{S}\left[y, \frac{l_a l_{b (a y)} l_{y z}}{l_{a b} l_y l_{(a y) z}}\right]+\mathcal{S}\left[(b-z), \frac{l_{b (a y)}}{l_{(a y)
   z}}\right]\\&+\mathcal{S}\left[z, \frac{l_b l_{y z} l_{a (b z)}}{l_{a b} l_z l_{y (b z)}}\right]+\mathcal{S}\left[l_a, \frac{a b y z l_{y z}}{l_{(a y) z}^2
   l_{a (b z)}}\right]+\mathcal{S}\left[l_b, \frac{a b y z l_{y z}}{l_{b (a y)} l_{y (b z)}^2}\right]+\mathcal{S}\left[l_{a b}, \frac{l_{b (a y)}^2 l_{a (b
   z)}^2}{a b y z l_a l_b l_y l_z}\right] \nonumber \\&+\mathcal{S}\left[D_1, \frac{l_b l_{y z}}{l_{a b} l_z}\right]+\mathcal{S}\left[D_2, \frac{l_a l_{y z}}{l_{a b}
   l_y}\right]+\mathcal{S}\left[D_5, \frac{l_{a b} l_y l_{(a y) z}}{l_a l_{b (a y)} l_{y z}}\right]+\mathcal{S}\left[D_{14}, \frac{l_{a b} l_z l_{y (b z)}}{l_b
   l_{y z} l_{a (b z)}}\right]\nonumber\\
&\nonumber\\
\mathcal{S}_{\mathcal{Z}_{44}}^{\text{II}}&=\mathcal{S}\left[a, \frac{l_a l_b}{l_{a b}}\right]+\mathcal{S}\left[b, \frac{l_{a b}}{l_a
   l_b}\right]+\mathcal{S}\left[(a-y), \frac{l_y}{l_a}\right]+\mathcal{S}\left[y, \frac{l_{y z}}{l_y
   l_z}\right]\\&+\mathcal{S}\left[(b-z), \frac{l_b}{l_z}\right]+\mathcal{S}\left[z, \frac{l_y l_z}{l_{y z}}\right]+\mathcal{S}\left[D_1, \frac{l_{a b} l_z}{l_b l_{y
   z}}\right]+\mathcal{S}\left[D_2, \frac{l_a l_{y z}}{l_{a b} l_y}\right]\nonumber\\
   &\nonumber\\
\mathcal{S}_{\mathcal{Z}_{44}}^{\text{III}}&=\mathcal{S}\left[b, \frac{l_a l_b}{l_{a b}}\right]+\mathcal{S}\left[(a-y), \frac{l_a}{l_y}\right]+\mathcal{S}\left[(b-z), \frac{l_{b (a y)} l_z}{l_b l_{(a y)
   z}}\right]+\mathcal{S}\left[z, \frac{l_{y z}}{l_y l_z}\right]+\mathcal{S}\left[l_a, \frac{l_{y z}}{l_{(a y) z}}\right]\\&+\mathcal{S}\left[l_{a b}, \frac{l_{b (a y)}}{l_b
   l_y}\right]+\mathcal{S}\left[l_y, \frac{l_{b (a y)}}{l_{a b}}\right]+\mathcal{S}\left[l_{y z}, \frac{l_a l_z}{l_{(a y) z}}\right]+\mathcal{S}\left[D_1, \frac{l_b l_{y
   z}}{l_{a b} l_z}\right]+\mathcal{S}\left[D_5, \frac{l_{a b} l_y l_{(a y) z}}{l_a l_{b (a y)} l_{y z}}\right]\nonumber\\
      &\nonumber\\
\mathcal{S}_{\mathcal{Z}_{44}}^{\text{IV}}&=\mathcal{S}\left[a, \frac{l_a l_b}{l_{a b}}\right]+\mathcal{S}\left[(a-y), \frac{l_y l_{a (b z)}}{l_a l_{y (b z)}}\right]+
\mathcal{S}\left[y, \frac{l_{y z}}{l_y   l_z}\right]+\mathcal{S}\left[(b-z), \frac{l_b}{l_z}\right]+\mathcal{S}\left[l_b, \frac{l_{y z}}{l_{y (b z)}}\right]\\&
+\mathcal{S}\left[l_{a b}, \frac{l_{a (b z)}}{l_a   l_z}\right]+\mathcal{S}\left[l_z, \frac{l_{a (b z)}}{l_{a b}}\right]+\mathcal{S}\left[l_{y z}, \frac{l_b l_y}{l_{y (b z)}}\right]+\mathcal{S}\left[D_2, \frac{l_a l_{y   z}}{l_{a b} l_y}\right]+\mathcal{S}\left[D_{14}, \frac{l_{a b} l_z l_{y (b z)}}{l_b l_{y z} l_{a (b z)}}\right]\nonumber\\
         &\nonumber\\
\mathcal{S}_{\mathcal{Z}_{44}}^{\text{V}}&=
\mathcal{S}\left[a, \frac{l_a l_{b (a y)} l_{y z}}{l_{a b} l_y l_{(a y) z}}\right]+\mathcal{S}\left[b, \frac{l_b l_{y z} l_{a (b z)}}{l_{a b} l_z l_{y
   (b z)}}\right]+\mathcal{S}\left[(a-y), \frac{l_{a (b z)}}{l_{y (b z)}}\right]+\mathcal{S}\left[y, \frac{l_a l_{b (a y)} l_{y z}}{l_{a b} l_y l_{(a y)
   z}}\right]\\&
   +\mathcal{S}\left[(b-z), \frac{l_{b (a y)}}{l_{(a y) z}}\right]+\mathcal{S}\left[z, \frac{l_b l_{y z} l_{a (b z)}}{l_{a b} l_z l_{y (b
   z)}}\right]+\mathcal{S}\left[l_a, \frac{a b y z}{l_{(a y) z} l_{a (b z)}}\right]+\mathcal{S}\left[l_b, \frac{a b y z}{l_{b (a y)} l_{y (b z)}}\right]\nonumber\\&+\mathcal{S}\left[l_{a   b}\otimes \frac{l_{b (a y)} l_{a (b z)}}{a b y z}\right]+\mathcal{S}\left[l_y\otimes \frac{l_{b (a y)} l_{y (b z)}}{a b y
   z}\right]+\mathcal{S}\left[l_z, \frac{l_{(a y) z} l_{a (b z)}}{a b y z}\right]+\mathcal{S}\left[l_{y z}, \frac{a b y z}{l_{(a y) z} l_{y (b
   z)}}\right]\nonumber\\&+\mathcal{S}\left[D_1, \frac{l_b l_{y z}}{l_{a b} l_z}\right]+\mathcal{S}\left[D_2, \frac{l_a l_{y z}}{l_{a b} l_y}\right]+\mathcal{S}\left[D_5, \frac{l_{a
   b} l_y l_{(a y) z}}{l_a l_{b (a y)} l_{y z}}\right]+\mathcal{S}\left[D_{14}, \frac{l_{a b} l_z l_{y (b z)}}{l_b l_{y z} l_{a (b z)}}\right]\nonumber
   \end{align}
\end{subequations}

\vskip 1.25 cm

\section{Conclusions}

We achieved a fully analytic evaluation of the one-loop five-point gluing process discussed in \cite{shotaThiago2,usFivePoints}. Our technique is as follows: as in the earlier attempts we close the integration contours for the particle rapidities in the complex plane and pick residues, thereby transforming the original sum-integral into a range of pure sums. Most of these are five-fold, but pulling a derivative onto the BES phase we create some six-fold cases. Even including the complicated bound state scattering matrix \cite{glebBound,ETH} these can be re-summed into Euler integrals. In the process, the scattering matrix is resolved into additional parametric integrals over a rational expression. 

Gluing is inherently finite so that most of our Euler integrals can be directly integrated to dilogarithms (or products of two simple logarithms). The transcendentality principle \cite{lipKot} is respected: any individual process will yield a result of logarithm weight two. However, some of the Euler integrals are multi-quadratic (or even cubic in one parameter) in which case we scale out some variables and use intersection theory for generalised hypergeometric functions \cite{inter1,inter2} to derive systems of Pfaffian equations on a basis of two-parameter master integrals. Choosing $dLog$ bases we obtain \emph{canonical} equations in the sense of \cite{hennAlg} that can easily be solved in terms of Goncharov logarithms. The Euler integrals that we submit to this procedure are finite, but they are not of uniform transcendentailty. We find terms of logarithm weights zero through two. These results are scaled back and the remaining parameters are integrated out. On the way we may meet trilogarithms, though like rational terms and simple logarithms they will cancel. 

We are clearly seeing the fundamentals of an integration technique for gluing processes. To attempt an analytic two-loop four-point or one-loop six-point computation is still a daunting prospect and would certainly require automation. The present attempt is guided by the experience with the simplest six-parameter series yielding multi-quadratic denominators \cite{usInter}. This effort pointed out how to use intersection theory for our purposes. Here we exploit that the same approach can be made to work for all elements of the bound state scattering matrix simply by variable rescalings: we can use one and the same intersection problem for all flavours of bound states. A similar road of attack seems recommendable for future studies of higher processes. Non-generic features of the current computation are that we singled out a special sequence of integrations for the Pfaffian equations in order to avoid integrating over quadratic arguments, and that we had to experiment eg. in Section \ref{hardest} as to how to combine the various blocks to avoid similar issues in the subsequent integration of the re-scaled intersection theory results.

As has been pointed out in \cite{shotaThiago1}, sum-integrals from gluing are akin to partially evaluated Mellin-Barnes representations \cite{MB} of Feynman integrals. And indeed, one of us has been able to exploit the methods developed in \cite{usFivePoints,usInter} and for the present article to re-analyse the off-shell and and two-loop box integrals by intersection theory \cite{meBox}. The Euler summability seems perfectly generic; further progress on the evaluation of Feynman integrals is certainly possible. There are very many interesting mathematical aspects of the related intersection theory programme. For instance, how does one single out the sequence of sums leading to the smallest Euler representation? How does one choose a $dLog$ basis so that all higher connections have only simple poles?

Next, at this point we can deal with multi-quadratic denominators in Euler integrals by home-grown methods. It should not be impossible to anaylse cubic or higher denominators introducing more technology. However, an efficient alternative might be \emph{parametric IBP} \cite{paraIBP} which has also recently be used in the context of \nFour SYM in \cite{smirnovRoman}.

Other domains in which the currently available methods are directly applicable are eg. POPE contributions \cite{POPE}. For example, the three series of residues relevant to the two-particle contribution in Regge kinematics studied in \cite{jamesGeorge} sum into Euler integrals with three, five and seven parameters, respectively, which are essentially multi-linear in the integration parameters. In fact, only the seven parameter case is quadratic in one of the parameters. By experience, if left to the end that integration will likely linearise, too. A complication is the high logarithm weight eight. Presumably, intersection theory is not needed to analytically integrate this problem; perhaps the methods of {\tt HyperInt} \cite{Panzer} can be adapted to the situation. Note that hypergeometric sums have originally come up in the context of the POPE in \cite{cordovaVanHippel}. We can claim to have pioneered iterating the process in \cite{usFivePoints}.

Third, finite size corrections in the thermodynamic Bethe ansatz (TBA) \cite{glebTBA} are a simpler version of those analysed here: they are very much the same integrals with the kinematic invariants put to 0, 1 or $\infty$. Assuming that the TBA yields single-valued $\zeta$ functions, can we introduce a pair of complex conjugate variables, evaluate via Pfaffian equations and then take a limit?

Last, upon re-summing into Euler integrals it is quite striking to see the bound state scattering matrix and the BES dressing phase dissolve into rational integrands. This is an interesting parallel to their disappearance from the quantum spectral curve \cite{QSC}. Perhaps a connection is given by the TBA?

\section*{Appendix A}

The $\cZ$ elements can also be expressed by a common factor $G_1(x_1^\mp, \, x_2^\mp)$ times a linear combination of only $\{ \cX^{k-1,l}_n, \, \cX^{k,l}_n \}$ and a second part similar to what is stated in \cite{ETH}, namely a linear combination of three $\cY$ matrices with shifted indices which the notebook then automatically decomposes into $\cX$ parts. For $\cZ_{33}$ we find the form
\beq
G_1 \eqsp -\frac{x_1^+-x_2^+}{g \, U_1 \, U_2 \, (x_1^+-x_2^-)(x_1^--x_2^+)(1-1/(x_1^+ x_2^+))} \, , \quad
G_2 \eqsp \frac{(x_1^+-x_1^-)(x_2^+-x_2^-)(1 - 1/(x_1^- x_2^-))}{U_1 \, U_2 \, (x_1^+-x_2^-)(x_1^--x_2^+)(1-1/(x_1^+ x_2^+))}
\eeq
and
\beq
R_1 \eqsp \, 2 i \, \{ 0, \, k+L-l, \, 0, \, \dupp - k + L -l \}
\eeq
while $R_2$ is as before, cf. \eqref{R12}. These formulae are to be decorated by the weight factor $W$, the measure $M$ and $D(x_1^\mp, \, x_2^\mp)$ from the $\cX$ matrices and rotated to $3 \gamma, \, 1\gamma$ kinematics. 

\newpage

It must be an advantage that there are only six blocks and that the coefficients in $R_1$ are simpler now. The computation is analogous to that of Section \ref{hardest} but there are some different features:
\begin{itemize}
\item Closing the integration contours, picking residues, and shifting the counters is done as above.
\item The $G_1 \, R_1 . \cX$ part decomposes into A type and B type terms under partial fractioning in $l$. We will take the sums in the sequence $\{l, \, L, \, j, \, k, \, K, \, m\}$ once again. However, it is useful to split the re-summation of this part into a full sum with $L \, > \, 0$ and a part with $L \eqsp 0$. Thus in a sense there are four blocks. None of this causes more than three integration parameters to appear.
\item There is a $k \eqsp 1, \, m \eqsp 0$ term to subtract for the $\cX^{k,l}_k$ part of the full sum. The sequence of the sums is somewhat subtle to establish here: to avoid creating too many parameters we may eg. try $\{l, \, K, \, L, \, j\}$ which yields a two-parameter and a three-parameter part. However, the latter three-parameter part will intermediately create trilogarithms upon integration. Fortunately, applying some care only two-parameter integrals occur with the sequence $\{K, \, l, \, L, \, j\}$.
\item We have to separate A and B type $j \eqsp 1$ terms from the six-fold sum. 
\item The five-fold $L \eqsp 0$ sum is fairly trivial using the sequences $\{l, \, k, \, m, \, K, \, j\}$ and $\{l, \, j, \, k, \, K, \, m\}$ in the A and B parts, respectively. In the sum of the resulting Euler integrals the B terms cancel whereas the A part yields a very simple three-parameter integral. 
\item In the B part of the $L \eqsp 0$ computation concerning $\cX^{k,l}_k$ we have to separately consider an $m \eqsp 0$ term arising in the $j$ step, which gives a simple two-parameter answer.
\item Here the $G_2 \, R_2 . \cX$ blocks all yield A type contributions only. We sum in the usual order $\{l, \, L, \, j, \, k, \, K, \, m\}$ keeping aside $j \eqsp 1$ terms. 
\item By Taylor expansion under the integral we can again check the completeness of the sum of all these parts. Its expansion does match the original series of residues from \cite{usFivePoints}. We now proceed as before, namely partially fractioning wrt. the parameter $s$ from the $L$ sum, checking the cancellation of divergences from poles in $1-s$, scaling (by the previous rules) and submitting  three-parameter B parts, and three- and four-parameter A parts to integral reduction by intersection theory, finally re-importing and integrating everything. The symbol of the result is identical to \eqref{prodZ33Sym}, \eqref{symS1S2}, \eqref{symS3S4}.
\end{itemize}
An obvious drawback of this version of the computation is the need to more seriously seek suitable sequences of summing.
Yet, the inconvenience is by far outweighed by a number of attractive features: in Section \ref{hardest} we had to engineer the trick of putting the A part of blocks \#1 and \#2 together to not run into square roots during the final integrations. Here we do not have to worry; the step is apparently preempted by the different form of $G_1, \, G_2, \, R_1$. Second, irreducible denominators naturally arise to maximally second power, ie. relation \eqref{shiftGamma} never needs to be invoked. In fact, third order poles do not occur at all so that the integral reduction is smoother.

Yet, the denominators of the B sector integrals contain the new polynomial
\beq
P_9 \eqsp a_0 - a_0 \, a \, \mathbf{q} - a \, b \, \mathbf{q} - a_0 \,  y + a_0 \, \mathbf{q} \, y - z + a \, \mathbf{q} \, z \label{defP9}
\eeq
beyond $P_1$ and $P_3$. We have set up a separate intersection theory computation for a family of integrals with these three denominators but without $P_4$. The basis size for this problem is 14. On the other hand, if $P_1, \, P_3, \, P_4, \, P_9$ are all allowed denominators in one big family of integrals one will need bases of length 16 for the two-parameter reduction. This is done in the bulk of our work to be able to rely on one common intersection theory framework.

\section*{Appendix B}

As stated in the main text, we need to enlarge the intersection theory problem discussed in Sections \ref{hardest}, \ref{bivariate} by the additional denominator polynomial \eqref{defP9} in order to handle all $S^v{\psi}$ in a common framework. The $u$ factor in \eqref{eq:u13} receives the additional term
\begin{equation}
\label{eq:twistInter}
u_{\text{complete}} \eqsp u \, P_9^{\gamma}
\end{equation}
and we can construct a basis $\langle e_i | = \hat{e}_i\, \text{d}\mathbf{s}\,\text{d}\mathbf{q}$ of 16 elements:
     \begin{align}
     \label{eq:basis16}
  \begin{split}
\hat{e}_i=
&\left\{\frac{1}{\mathbf{s} \, \mathbf{q} },\quad \frac{-1}{\mathbf{s} (1-\mathbf{q} )},\quad \frac{-1}{(1-\mathbf{s}) \mathbf{q} }, \quad\frac{1}{(1-\mathbf{s}) (1-\mathbf{q} )},\right. \frac{1}{\mathbf{s}\, P_3},\quad \frac{1}{\mathbf{s}\, P_9},\quad \frac{1}{\mathbf{s}\, P_4},\quad\frac{-1}{(1-\mathbf{s}) P_3}, \\&\quad\frac{1}{(1-\mathbf{s}) P_9},\quad\frac{1}{(1-\mathbf{s}) P_4},\frac{1}{P_1},\quad \frac{(1-\mathbf{q} )}{ P_1},\frac{1}{(1-\mathbf{s}) P_1}, \quad\frac{1}{\mathbf{s} P_1},\left. \frac{ (1-\mathbf{q} )}{P_1}, \quad\frac{ (1-\mathbf{q} )}{P_1 P_9}\right\}.
\end{split}
  \end{align}
  Following the steps explained in section \ref{bivariate} we find the 16 by 16 intersection matrix	
    \begin{equation}
  \text{\textbf{C}}=\frac{1}{84 \gamma^{2}}D^{-1}\hat{C}D^{-1}
  \end{equation}
  
  \begin{equation}
\hat{C}=\left(
\begin{array}{cccccccccccccccc}
 54 & 9 & 18 & 3 & 9 & 9 & 9 & -3 & 3 & 3 & 0 & 0 & 0 & 0 & 0 & 0 \\
 9 & 12 & 3 & 60 & -9 & -9 & -9 & 3 & -3 & -3 & 0 & 0 & 0 & 0 & 0 & 0 \\
 18 & 3 & 54 & 9 & 3 & 3 & 3 & -9 & 9 & 9 & 0 & 0 & 0 & 0 & 0 & 0 \\
 3 & 60 & 9 & 12 & -3 & -3 & -3 & 9 & -9 & -9 & 0 & 0 & 0 & 0 & 0 & 0 \\
 9 & -9 & 3 & -3 & 54 & -9 & -9 & -18 & -3 & -3 & 0 & 0 & 0 & 0 & 0 & 0 \\
 9 & -9 & 3 & -3 & -9 & 54 & -9 & 3 & 18 & -3 & 0 & 0 & 0 & 0 & 0 & 0 \\
 9 & -9 & 3 & -3 & -9 & -9 & 54 & 3 & -3 & 18 & 0 & 0 & 0 & 0 & 0 & 0 \\
 -3 & 3 & -9 & 9 & -18 & 3 & 3 & 26 & 9 & 9 & 0 & 0 & 28 & 0 & 0 & 0 \\
 3 & -3 & 9 & -9 & -3 & 18 & -3 & 9 & 54 & -9 & 0 & 0 & 0 & 0 & 0 & 0 \\
 3 & -3 & 9 & -9 & -3 & -3 & 18 & 9 & -9 & 26 & 0 & 0 & 28 & 0 & 0 & 0 \\
 0 & 0 & 0 & 0 & 0 & 0 & 0 & 0 & 0 & 0 & -168 & 0 & 0 & 0 & 0 & 0 \\
 0 & 0 & 0 & 0 & 0 & 0 & 0 & 0 & 0 & 0 & 0 & 168 & 0 & 0 & 0 & 0 \\
 0 & 0 & 0 & 0 & 0 & 0 & 0 & 28 & 0 & 28 & 0 & 0 & 112 & 0 & 0 & 0 \\
 0 & 0 & 0 & 0 & 0 & 0 & 0 & 0 & 0 & 0 & 0 & 0 & 0 & 168 & 0 & 0 \\
 0 & 0 & 0 & 0 & 0 & 0 & 0 & 0 & 0 & 0 & 0 & 0 & 0 & 0 & -24 & 0 \\
 0 & 0 & 0 & 0 & 0 & 0 & 0 & 0 & 0 & 0 & 0 & 0 & 0 & 0 & 0 & 168 \\
\end{array}
\right)
  \end{equation}
  
  with
  
  \begin{equation}
  \label{eq:diagonalAndDenominator}
  D_{\mathbf{q}}=\text{diag}\left(1,\,1,\,1,\,1,\,a-y,\,D_{12},\,D_1,\,a-y,\,D_{12},\,D_1,\, y\sqrt{D_9},\, y D_6,\,D_4,\,D_7,\, y D_8,\, y D_{13}\right).
  \end{equation}
 
 Here we defined the additional polynomials:
\begin{align}
&D_{12} \eqsp a \, b-a_0 \, y-a \, z+a_0 \, a \, , \notag \\
&D_{13} \eqsp a \, a_0 \, b - a \, a_0^2 \, b - a \, a_0 \, b^2 - a \, a_0 \, b \, y + a \, a_0^2 \, b \, y + a \, a_0 \, b^2 \, y +  a \, a_0 \, b \, z \\
&\quad\quad+ a \, b^2 \, z - a_0 \, y \, z + a \, a_0 \, y \, z + a_0^2 \, y \, z - a \, a_0^2 \, y \, z - a \, a_0 \, b \, y \, z - a \, b \, z^2. \notag
 \end{align}

 \section*{Appendix C}

We include five ancillary files: {\tt intsB3A3A4.m} contains the two three-parameter B sector integrals, the two three-parameter A sector integrals, and the two four-parameter A sector integrals, collected in the array {\tt intsS}. The result of the integral reduction is given in {\tt outZ33IntsB3A3A4.m}. In the output we have added the two integrals of each type. The variables {\tt const1, const2, const3} contain the respective rational parts,  {\tt finLog1, finLog2, finLog3} give the coefficients of single logarithms in the basis {\tt logBasFV}. Finally, {\tt facPoly2, facPoly3} give the coefficients of the dilogarithm function {\tt polyFunc} for the A sector three- and four-parameter integrals, respectively.

The files {\tt AppAB3A3A4.m} and {\tt outZ33AppAB3A3A4.m} contain the same variables for the second $\cZ_{33}$ computation along the lines of Appendix A. Note that the dilogarithm part is identical.

The concise file {\tt  polyGammaZ33.m} gives the symbol for the contribution of the polygamma residue to scattering by $\cZ_{33}$ as written up in Section \ref{hardest} and the corresponding four polylogarithm functions. Finally, in {\tt blocksZ33.m} the variable {\tt all331} contains the eight six-fold sums of Section \ref{hardest} prior to re-summation, and {\tt all331K1M0} a subtraction term.

\end{document}